\documentclass[journal]{IEEEtranTIE}
\usepackage{graphicx}
\usepackage{cite}
\usepackage{picinpar}
\usepackage{amsmath}
\usepackage{url}
\usepackage{flushend}
\usepackage[utf8]{inputenc}
\usepackage{colortbl}
\usepackage{soul}
\usepackage{multirow}
\usepackage{pifont}
\usepackage{color}
\usepackage{alltt}
\usepackage[hidelinks]{hyperref}
\usepackage{enumerate}
\usepackage{siunitx}
\usepackage{breakurl}
\usepackage{epstopdf}
\usepackage{pbox}

\usepackage[caption=false,font=footnotesize]{subfig}

\begin{document}
\title{Decoupled Internal Energy Regulation and Inertial Response Provision for Grid-Forming Multilevel-Converter-Based E-STATCOMs}

\author{
	\vskip 1em
	
	Ki-Hyun Kim, 
	Yeongung Kim, \emph{Graduate Student Member, IEEE},
	Shenghui Cui,
	and Jae-Jung Jung, \emph{Senior Member, IEEE}

	\thanks{
		(\emph{corresponding author: Jae-Jung Jung and Shenghui Cui})

		Ki-Hyun Kim, Yeongung Kim and Jae-Jung Jung are with the School of Electronic and Electrical Engineering, 
		Kyungpook National University, Daegu, 41566, South Korea (e-mail: rlarlgus5615@knu.ac.kr; free77a@knu.ac.kr; jj.jung@knu.ac.kr). 
		
		Shenghui Cui is with Department of Electrical and Computer Engineering, 
		Seoul National University, Seoul, 08826, South Korea (e-mail: cuish@snu.ac.kr)
	}
}

\maketitle
	
\begin{abstract}
As power systems accommodate higher shares of renewable generation, short-term power imbalances become more frequent and can manifest as pronounced voltage and frequency excursions in low-inertia conditions. 
E-STATCOMs—STATCOMs equipped with energy storage—offer a practical means to provide both voltage support and fast frequency assistance under grid-forming control. 
Among candidate implementations, double star multilevel-converter (DS-MC) -based E-STATCOMs enable centralized energy-storage integration at the dc-link, which improves thermal management and maintainability. 
Nevertheless, conventional dc-side power based internal-energy regulation in DS-MCs can undesirably couple loss compensation to the energy-storage path, accelerating storage cycling and constraining operation when the storage is unavailable. 
This paper introduces a control strategy that assigns DS-MC total internal-energy regulation to the ac-side active power path, while reserving the dc-side storage power solely for frequency support. 
By decoupling internal-energy management from inertial-response provision, the proposed scheme enables flexible operation as either a STATCOM or an E-STATCOM according to storage availability and mitigates unnecessary storage cycling. 
The proposed strategy is verified through offline simulations and laboratory-scale experiments.
\end{abstract}


\markboth{IEEE TRANSACTIONS ON INDUSTRIAL ELECTRONICS}%
{}

\definecolor{limegreen}{rgb}{0.2, 0.8, 0.2}
\definecolor{forestgreen}{rgb}{0.13, 0.55, 0.13}
\definecolor{greenhtml}{rgb}{0.0, 0.5, 0.0}

\section{Introduction}

\IEEEPARstart{A}{s} the penetration of renewable energy sources increases, renewable intermittency and limited predictability exacerbate short-term mismatches between power supply and demand. 
In conventional synchronous-generator-based systems, synchronous-machine rotational inertia provides an inherent energy buffer that mitigates such mismatches. 
However, power-electronic-based resources operating under grid-following (current-controlled) control do not contribute comparable inertial energy. 
Consequently, the resulting imbalance can induce voltage-magnitude and frequency fluctuations and, under severe conditions, may ultimately compromise system stability \cite{1}, \cite{2}.
To address these challenges, the energy-storage-enhanced static synchronous compensator (E-STATCOM), which integrates a static synchronous compensator (STATCOM) with an energy storage (ES), such as batteries or supercapacitors, has emerged as a promising grid-stabilization solution. 
Beyond the conventional STATCOM functionalities of voltage regulation and reactive power compensation at the point of common coupling (PCC), an E-STATCOM can adopt grid-forming (GFM) control to emulate inertial response and support frequency dynamics. 
Moreover, depending on the power and energy capabilities of the integrated storage, the E-STATCOM can provide sustained active power injection, thereby enabling fast primary frequency support. 
Consequently, E-STATCOMs can improve power quality and strengthen both voltage and frequency stability in modern power systems \cite{3}, \cite{4}, \cite{5}.

\begin{figure}[!t]
	\centering
	\subfloat[]{\includegraphics[width=4.25cm]{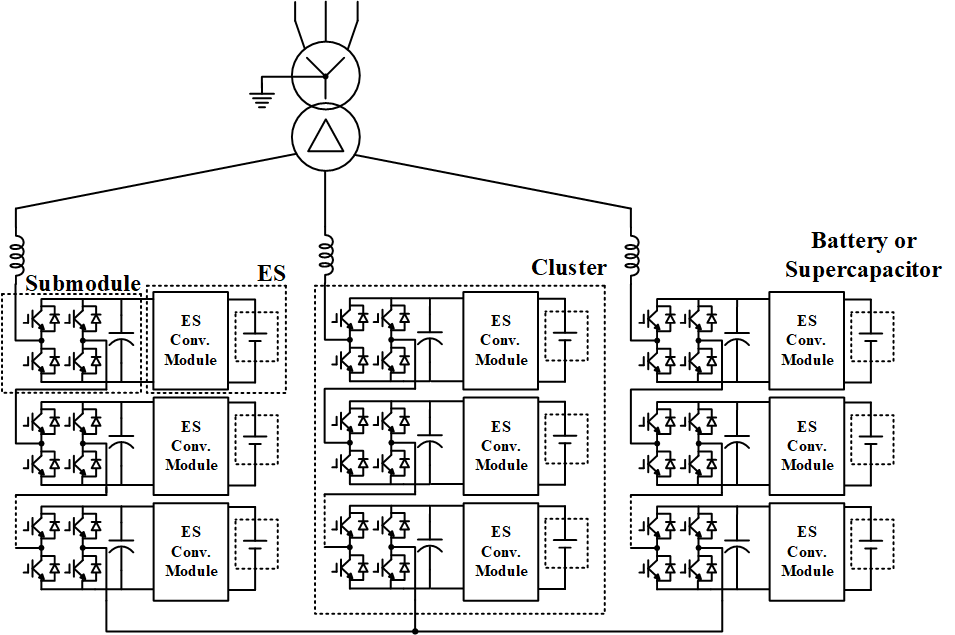}\label{fig:1a}}
	\subfloat[]{\includegraphics[width=4.25cm]{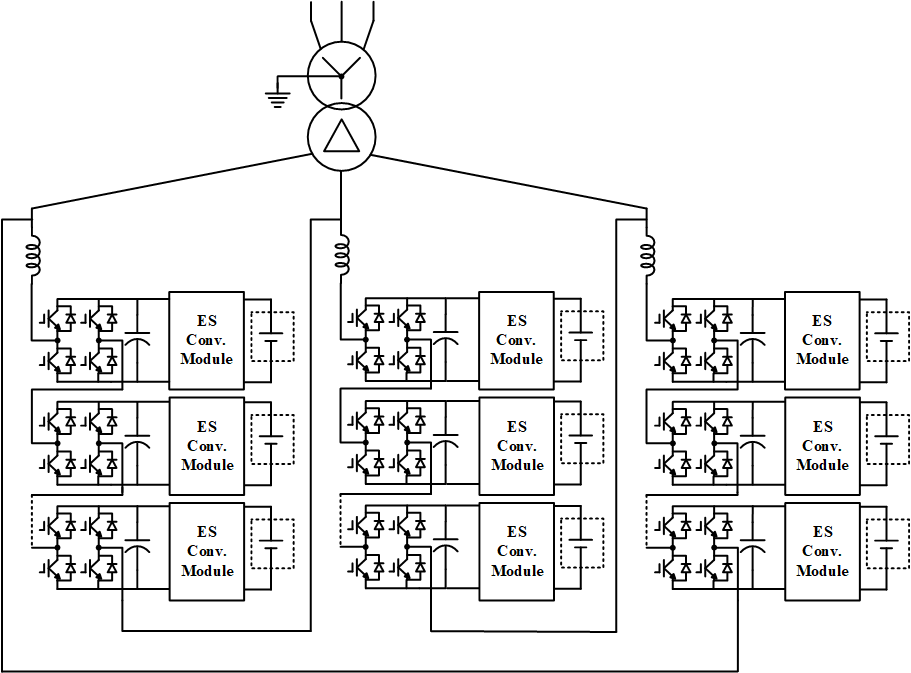}\label{fig:1b}}\\
	\subfloat[]{\includegraphics[width=4.25cm]{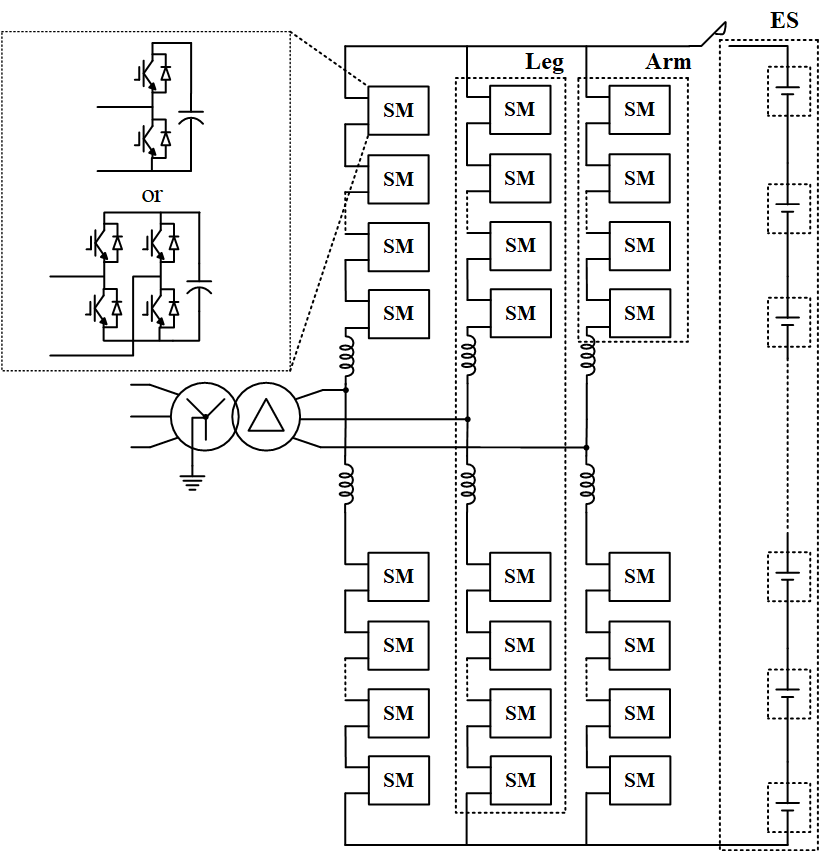}\label{fig:1c}}
	\caption{E-STATCOM topologies. (a) E-STATCOM based on single-star multilevel converter. (b) E-STATCOM based on single-delta multilevelconverter. (c) E-STATCOM based on double-star multilevel converter.}\label{FIG_1}
\end{figure}

E-STATCOM topologies reported in the literature predominantly adopt multilevel converters \cite{6}, \cite{7}, \cite{8}.
As illustrated in Fig. \ref{FIG_1}, multilevel converters are composed of a series connection of submodules, each consisting of power semiconductor switches and a dc-link capacitor. 
Owing to this modular structure, multilevel converters offer scalable voltage levels and are well suited for high-voltage and high-power applications \cite{9}. 
In addition, multilevel converters can achieve low output-current total harmonic distortion (THD) and high-quality output-voltage waveforms even at low switching frequencies, while providing redundancy against submodule failures. 
These advantages make multilevel converters a particularly promising topology for E-STATCOM applications.

As illustrated in Fig. \ref{FIG_1}\subref{fig:1a} and \subref{fig:1b}, E-STATCOMs implemented with single-star and single-delta multilevel converters typically adopt a distributed ES arrangement because a common dc-link is not available. 
The distributed ES is typically co-located with the power stage at the submodule level, and the resulting thermal proximity to loss-generating power semiconductor switches can expose the ES to harsh temperature conditions, potentially exceeding the recommended operating range \cite{10}, \cite{11}. 
Such thermal-management issues can accelerate ES aging, reduce lifetime, and increase both cost and cooling-system complexity. Moreover, even though redundancy can maintain operation after an ES fault, shutdown of the entire E-STATCOM is often required to safely service and replace the faulty ES unit \cite{10}.

By contrast, as shown in Fig. \ref{FIG_1}\subref{fig:1c}, double-star multilevel-converter (DS-MC) based E-STATCOMs allow a centralized ES to be installed at the dc-link \cite{12}, \cite{13}. 
This configuration decouples ES placement from the power stage, enabling independent installation and tighter temperature regulation, which extends ES lifetime and simplifies system management \cite{14}. 
In addition, dc-link isolation using disconnectors permits the E-STATCOM to continue operating as a conventional STATCOM while ES maintenance is performed \cite{4}, \cite{10}. 
Accordingly, Fig. \ref{FIG_1}\subref{fig:1c} has attracted increasing interest as a practical DS-MC-based E-STATCOM topology.

Stable operation of a DS-MC requires regulation of the total internal energy. \cite{15}
The total internal energy of a DS-MC can be regulated through either the dc-side power exchange or the ac-side active power exchange. Under GFM operation, when total internal energy regulation is assigned to the ac side, two practical implementations are commonly used. 
The first employs a cascaded structure, where the total-energy controller (TEC) generates an active power reference and the active power controller (APC) generates the converter frequency and angle commands from the active power error. In such a cascaded configuration, stable operation requires a clear separation of control bandwidths \cite{16}, \cite{17}.
If the TEC is tuned too slowly, rapid stabilization of the total internal energy becomes challenging. 
Conversely, if the APC is tuned too fast, the converter may no longer exhibit the desired GFM capabilities, such as providing inertial response. 
The second adopts a direct energy-synchronization control, in which the TEC directly generates the frequency deviation from the total-energy error. 
However, existing studies indicate that this approach is prone to low-frequency oscillatory instability \cite{18}. 

On the other hand, when the dc-side power is used to maintain the total internal energy of the DS-MC, the ES continuously participates in compensating the internal losses of the DS-MC. 
As a result, the ES is subjected to persistent charge-discharge cycling, which can increase degradation and maintenance requirements over the lifetime of ES. 
In addition, when the total internal energy is regulated through the dc-side power, the E-STATCOM cannot operate in the conventional STATCOM mode when the ES is unavailable.

Therefore, this paper proposes a novel control strategy that regulates the total internal energy of the DS-MC through the ac-side active power, thereby minimizing unnecessary ES charge-discharge cycling. 
Accordingly, the dc-side power—provided by the ES—is reserved exclusively for frequency response rather than for compensating the DS-MC internal losses. 
In addition, the proposed strategy decouples the internal-energy dynamics of the DS-MC from the provision of inertial energy, enabling flexible operation as either a STATCOM or an E-STATCOM depending on ES availability.

\section{Operating Schemes of DS-MC based E-STATCOMs}

\subsection{System configuration}

Fig. \ref{FIG_2} illustrates an arm-averaged model of the three-phase DS-MC, where the submodule capacitor voltages are assumed to be balanced and represented by an equivalent arm capacitance \(C\).
\(V_{dc}\) denotes the dc-link voltage.
\(v_{au}\), \(i_{au}\) and \(m_{au}\) denote the upper-arm voltage, current, and modulation index of phase \(a\), respectively, whereas \(v_{al}\), \(i_{al}\), and \(m_{al}\) denote the corresponding lower-arm quantities. 
In addition, \(i_{as}\) denotes the ac-side current of phase \(a\). 
\(R_{arm}\) and \(L_{arm}\) represent the arm resistor and arm inductor, respectively.

The DS-MC consists of six arms, each formed by a series connection of submodules implemented as either half-bridge or full-bridge cells.
However, for E-STATCOM applications, full-bridge-cell-based submodules are preferred, as they provide fault ride-through (FRT) capability and enable a significantly wider dc-link voltage operating range \cite{12}, \cite{13}.
In each phase, the upper and lower arms constitute one leg, and the three legs form the three-phase DS-MC.
For stable DS-MC operation, the energies of the six arms must be properly balanced; this section assumes that arm-energy balancing is achieved.

\subsection{Operating schemes for total internal energy regulation in DS-MC-based E-STATCOMs}

\begin{figure}[!t]
	\centering
	\subfloat[]{\includegraphics[width=4.25cm]{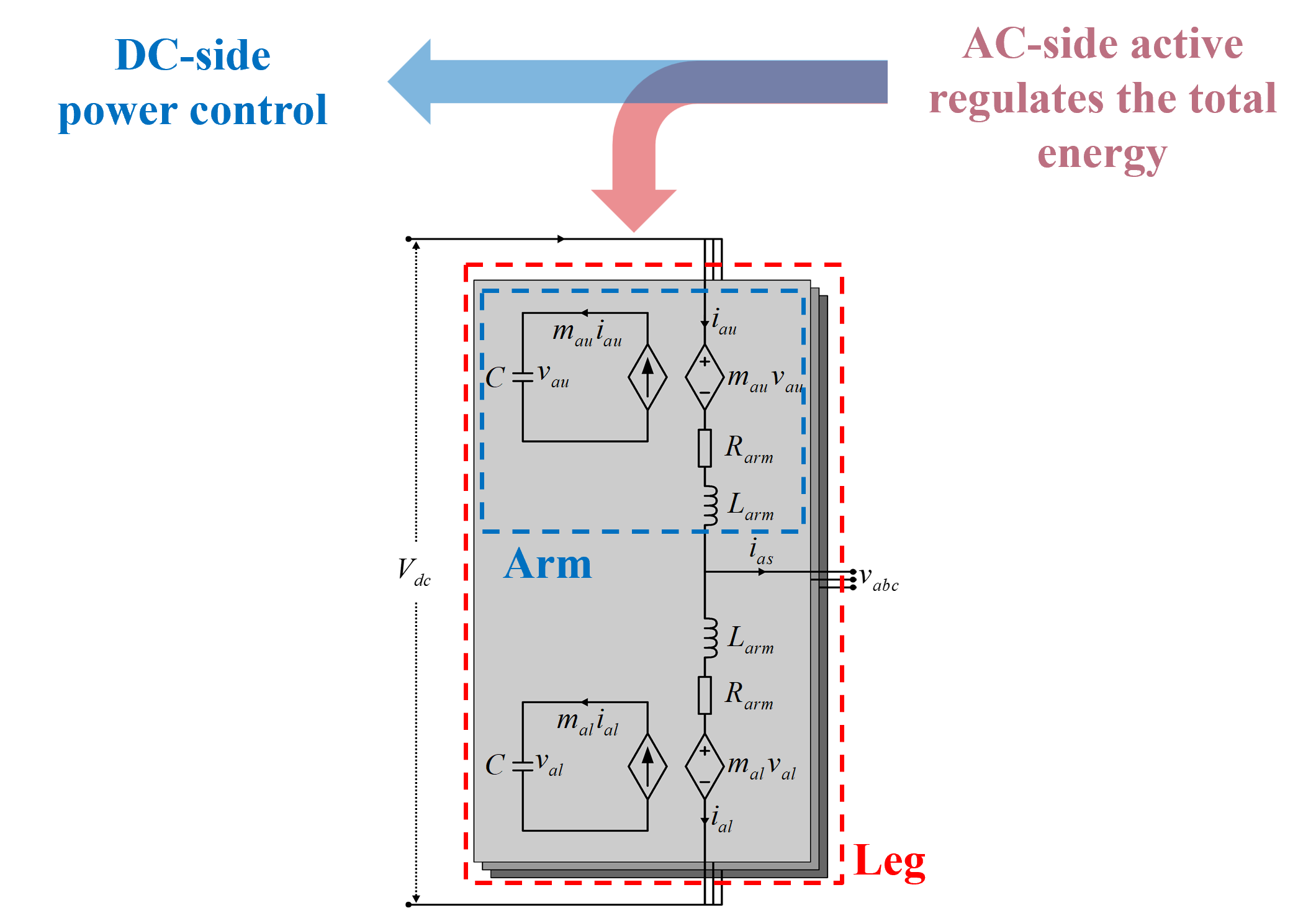}\label{fig:2a}}
	\subfloat[]{\includegraphics[width=4.25cm]{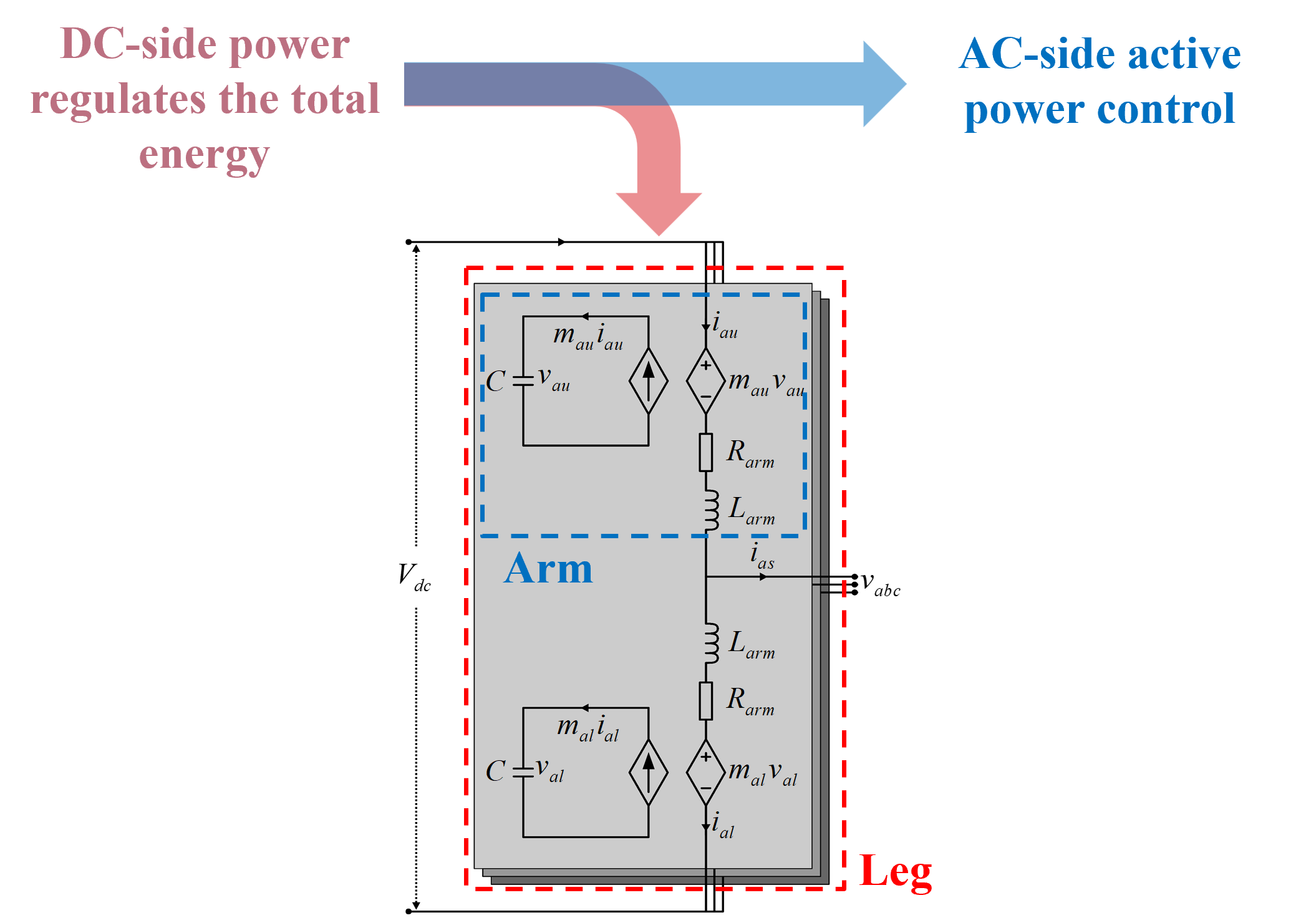}\label{fig:2b}}
	\caption{Operating schemes for total internal-energy regulation in a DS-MC-based E-STATCOM. (a) Scheme I: Total internal-energy regulation via ac-side active power, while dc-side power is used for power control. (b) Scheme II: Total internal-energy regulation via dc-side power, while ac-side active power is used for power control.}\label{FIG_2}
\end{figure}

Unlike two-level converters, DS-MCs decouple the arm-capacitor energy dynamics from the dc-link voltage, thereby providing two effective control degrees of freedom.
This property enables independent regulation of the ac-side power via the differential-mode arm-voltage component (i.e., \((v_{al}-v_{au})/2\)), whereas the common-mode component (i.e., \((v_{au}+v_{al})/2\)) governs the dc-side power exchange.
Beyond arm-energy balancing, stable DS-MC operation also requires regulation of the total internal energy, defined as the sum of the energies stored in the six arms.
In steady state, maintaining the total internal energy implies that the ac-side active power and dc-side power are properly balanced. 
Hence, one degree of freedom must be allocated to internal energy regulation, which can be realized via either the ac-side active power (Fig. \ref{FIG_2}\subref{fig:1a}, hereafter referred to as scheme I) or the dc-side power (Fig. \ref{FIG_2}\subref{fig:1b}, hereafter referred to as scheme II), while the remaining degree of freedom is used for the outer-loop active power regulation objective.

\subsection{Challenges of operating schemes under GFM control}

Scheme I generally employs a cascaded structure in which the TEC generates an active power reference, while the APC produces the converter frequency and reference angle based on the active power error.
In this configuration, the two closed loops are inherently coupled through the same ac-side active power path. 
As a result, stable operation requires a clear separation of control bandwidths \cite{16}, \cite{17}: the total energy loop must be tuned sufficiently slower than the APC loop so that internal-energy regulation does not interfere with the frequency/angle generation dynamics.
This requirement inevitably constrains the achievable bandwidth of total internal energy regulation, making rapid stabilization of the DS-MC internal energy challenging during transients.

\begin{figure}[!t]\centering
	\includegraphics[width=8.5cm]{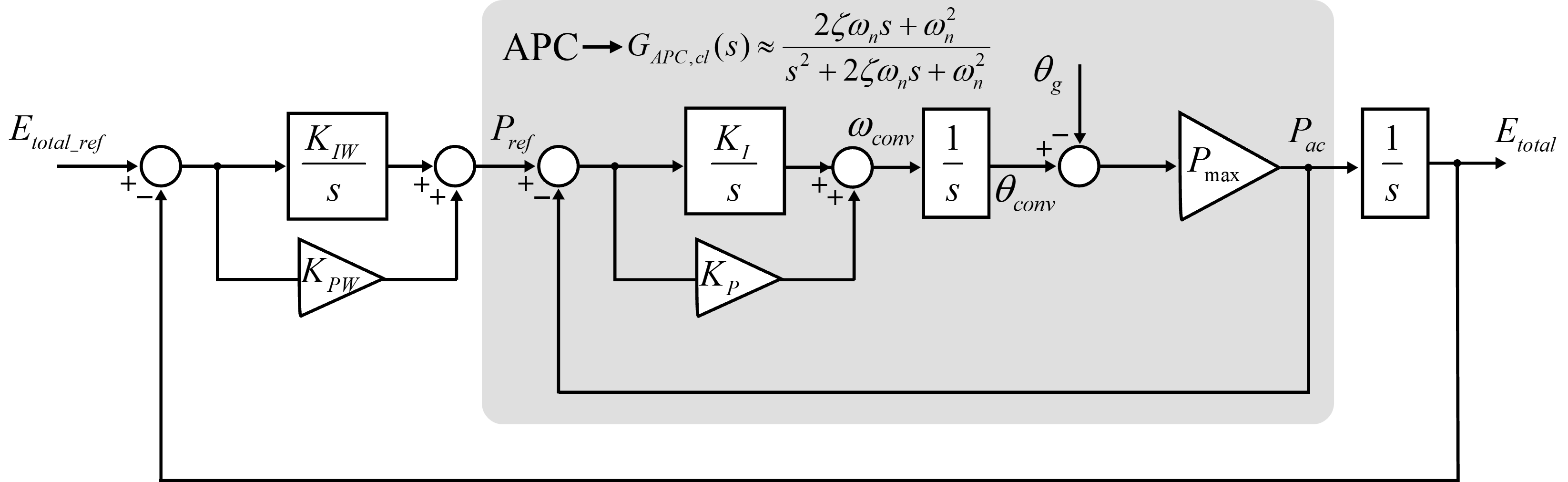}
	\caption{Small-signal closed-loop block diagram of the cascaded TEC-APC structure.}\label{FIG_3}
\end{figure}

Fig. \ref{FIG_3} shows the small-signal closed-loop block diagram of the cascaded TEC-APC structure.
Both controllers employ proportional-integral (PI) controllers, where the subscript \(W\) denotes the gains of the TEC.
Let \(E\) and \(V\) denote the DS-MC internal voltage source and the grid voltage, respectively, and let \(X\) denote the equivalent reactance between the two voltages.
Then, \(P_{max}\) is given by \(EV/X\).
In \cite{19}, the closed-loop transfer function of a PI-controller-based APC can be approximated by a second-order low-pass filter with a left-half-plane zero, as expressed in (1).
Accordingly, the loop gain in Fig. 3 is obtained as (2).

\begin{align}
G_{\mathrm{APC,cl}}(s) \approx \frac{2\zeta\omega_n s + \omega_n^2}{s^2 + 2\zeta\omega_n s + \omega_n^2} .
\end{align}

\begin{align}
G_{op}(s) = (K_{PW} + \frac{K_{IW}}{s}) \cdot G_{\mathrm{APC,cl}}(s) \cdot \frac{1}{s}
\end{align}

\begin{figure}[!t]\centering
	\includegraphics[width=8.5cm]{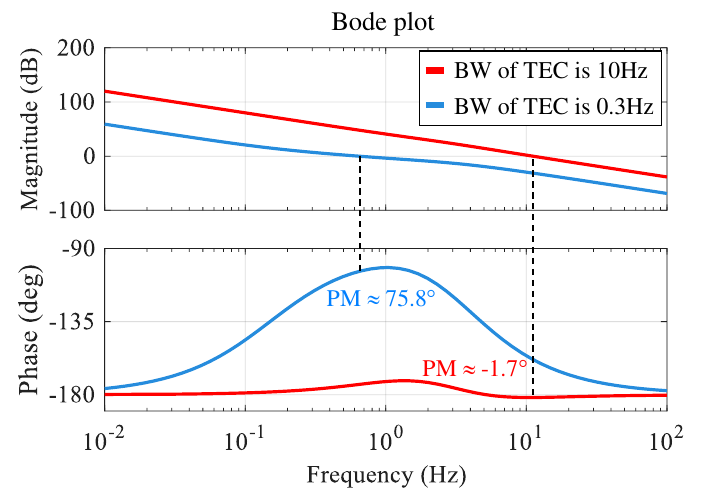}
	\caption{Bode plots for loop gain of the cascaded TEC-APC structure with the APC bandwidth fixed at 3 Hz.}\label{FIG_4}
\end{figure}

Fig. \ref{FIG_4} presents the bode plots of the loop gain for two cases in which the APC bandwidth is fixed at 3 Hz, while the bandwidth of the TEC is set to 10 Hz and 0.3 Hz, respectively.
With the TEC bandwidth of 0.3 Hz, the cascaded TEC-APC system secures a sufficient pahse margin of approximately 75.8\(^\circ\).
However, when the TEC bandwidth increases to 10 Hz, the phase margin drops to approximately - 1.7\(^\circ\), rendering the system unstable.
The results confirm that insufficient bandwidth separation reduces the phase margin and may render the overall closed-loop system unstable.

Moreover, excessively increasing the APC bandwidth to improve internal-energy regulation may compromise the intended GFM behavior by eliminating the “slow voltage-source” dynamics that enable fast active/reactive power responses to grid disturbances.
This, in turn, undermines the motivation for adopting GFM control in E-STATCOM applications.
The need to limit the APC bandwidth is also reflected in grid-code recommendations (e.g., \cite{20}, \cite{21}), which impose bandwidth constraints on the synchronization loop of GFM resources.

In Scheme II, the total internal energy is regulated through the dc-side power provided by the ES, while the ac-side degree of freedom is used for the active power control.
Under this allocation, the ES must continuously compensate the internal losses of DS-MCs in steady state to maintain the total internal energy. 
Consequently, the ES is subjected to persistent charge-discharge cycling even in the absence of frequency-support events, which increases RMS current and thermal stress, potentially accelerating ES degradation and raising maintenance requirements.
In addition, because total-internal energy regulation relies on the dc-side power, the E-STATCOM cannot seamlessly fall back to the conventional STATCOM mode when the ES is unavailable (e.g., during faults, disconnection, or maintenance), thereby reducing operational flexibility.

Therefore, to address the limitations of the two schemes, the next section proposes an E-STATCOM control strategy under GFM operation that stabilizes the total internal energy through the ac-side active power while reserving the ES exclusively for frequency response.

\section{Proposed GFM Control Strategy for Decoupling Inertial Response and Internal-Energy Regulation}

\subsection{System Configuration and ES Selection of the proposed control strategy}

In DS-MC-based E-STATCOMs, the ES is installed at the dc-link to provide fast active-power support under GFM operation.
The ES can be realized using batteries or supercapacitors.
In this study, a supercapacitor-based ES is adopted due to its high power capability and long cycling life, despite its relatively limited energy density \cite{22}.
In addition, to minimize the required supercapacitor capacity, the proposed strategy dedicates the ES power exclusively for inertial response, rather than for primary frequency control \cite{23}.

\begin{figure}[!t]\centering
	\includegraphics[width=8.5cm]{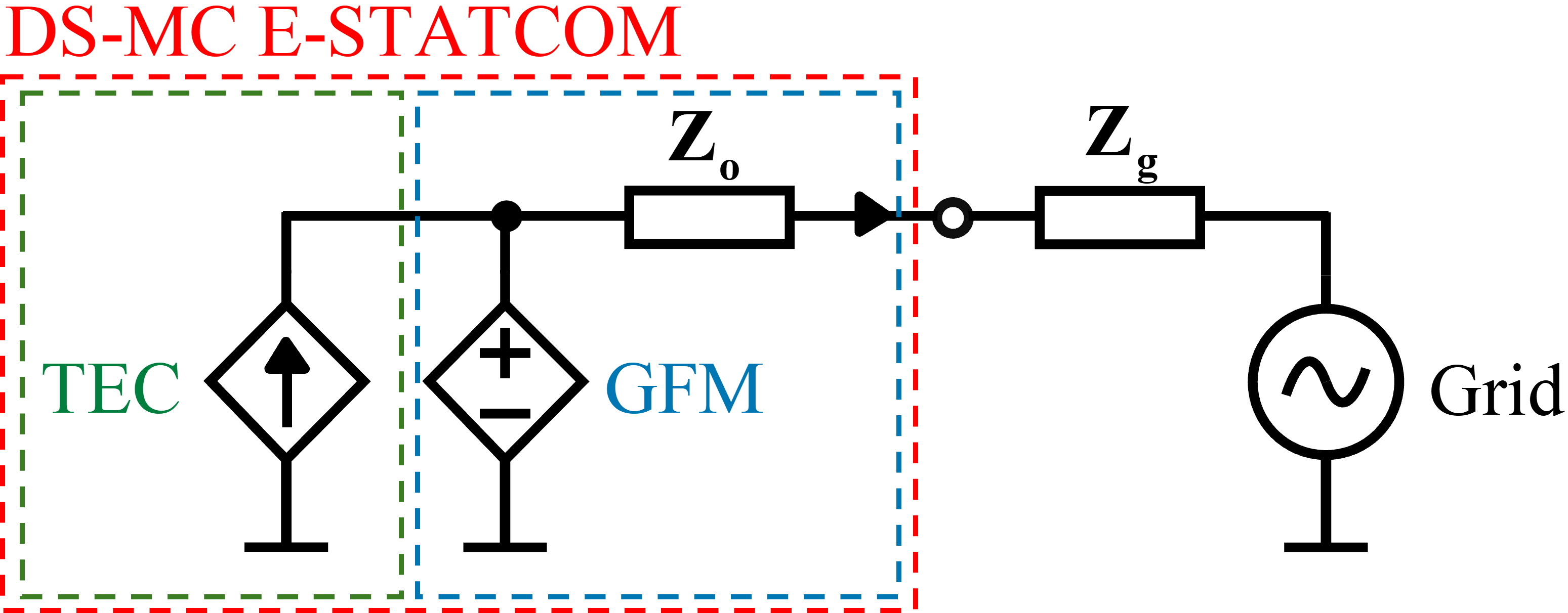}
	\caption{Equivalent circuit of the proposed E-STATCOM control strategy.}\label{FIG_5}
\end{figure}

Fig. \ref{FIG_5} presents a simplified equivalent circuit of the proposed DS-MC based E-STATCOM control strategy. 
Owing to the decoupling between total internal energy regulation (via the TEC) and inertial energy provision (via the APC), the TEC is modeled as a current source, whereas the inertial-response contribution is represented as a voltage source.

\begin{figure}[!t]\centering
	\includegraphics[width=8.5cm]{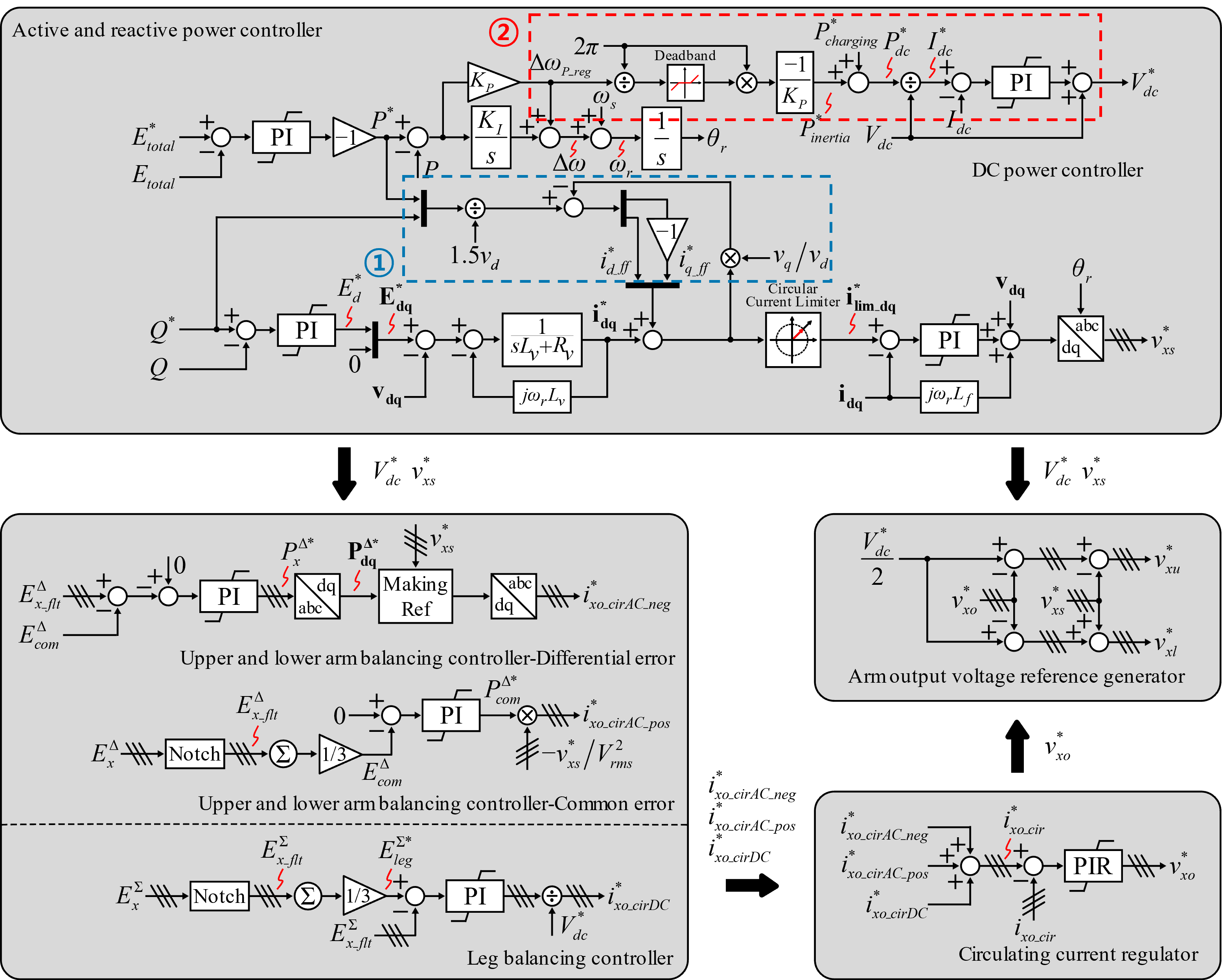}
	\caption{Overall control block diagram of the proposed control strategy.}\label{FIG_6}
\end{figure}

Fig. \ref{FIG_6} illustrates the overall control block diagram of the proposed control strategy for the DS-MC based E-STATCOM.
For the GFM operation, the outer loops employ the PI-controller-based APC and reactive power controller (RPC), while the inner loop adopts virtual admittance control with a current controller.
In addition, a balancing controller and a circulating-current controller are included to ensure arm-energy balancing of the DS-MC.
Furthermore, the proposed control framework incorporates a current-reference feedforward term (blue dotted box) to realize current-source behavior of the TEC, as well as an inertial power reference mapping term (red dotted box) to coordinate the ES inertial response contribution through the dc current controller. 

\subsection{Derivation of the Current-Reference Feedforward Term}

In the cascaded TEC-APC structure, bandwidth separation becomes critical because a slow APC cannot effectively suppress the large active-power error induced by fast TEC dynamics.
Accordingly, by enforcing current-source behavior in the TEC to rapidly shape the active-power feedback, the APC can regulate the remaining small active-power error effectively even with a slow dynamic characteristic.
This is because both total internal-energy regulation and inertial response are realized through the same PCC active-power degree of freedom; hence, their contributions appear superimposed in the PCC active power.
Consequently, by computing the current-reference feedforward term from the output of the TEC, as shown in (3)-(4), the total-energy-regulation component—excluding the inertial-response contribution—can be made to behave as a current source.
\(P_{ref}\) denotes the output of the TEC, and the \(P_{ac}\) denotes the active power at the PCC.
In addition, the subscripts \(d\) and \(q\) represent the \(d-\) and \(q-\) axis components in the converter synchronous reference frame, respectively.

\begin{align}
P_{ac} = 1.5(v_{PCC\_d}i_{PCC\_d} + v_{PCC\_q}i_{PCC\_q}) .
\end{align}

\begin{align}
i_{dref\_ff} = \frac{P_{ref}}{1.5v_{PCC\_d}} - \frac{v_{PCC\_q}}{v_{PCC\_d}}i_{qref} .
\end{align}

\begin{figure}[!t]\centering
	\includegraphics[width=8.5cm]{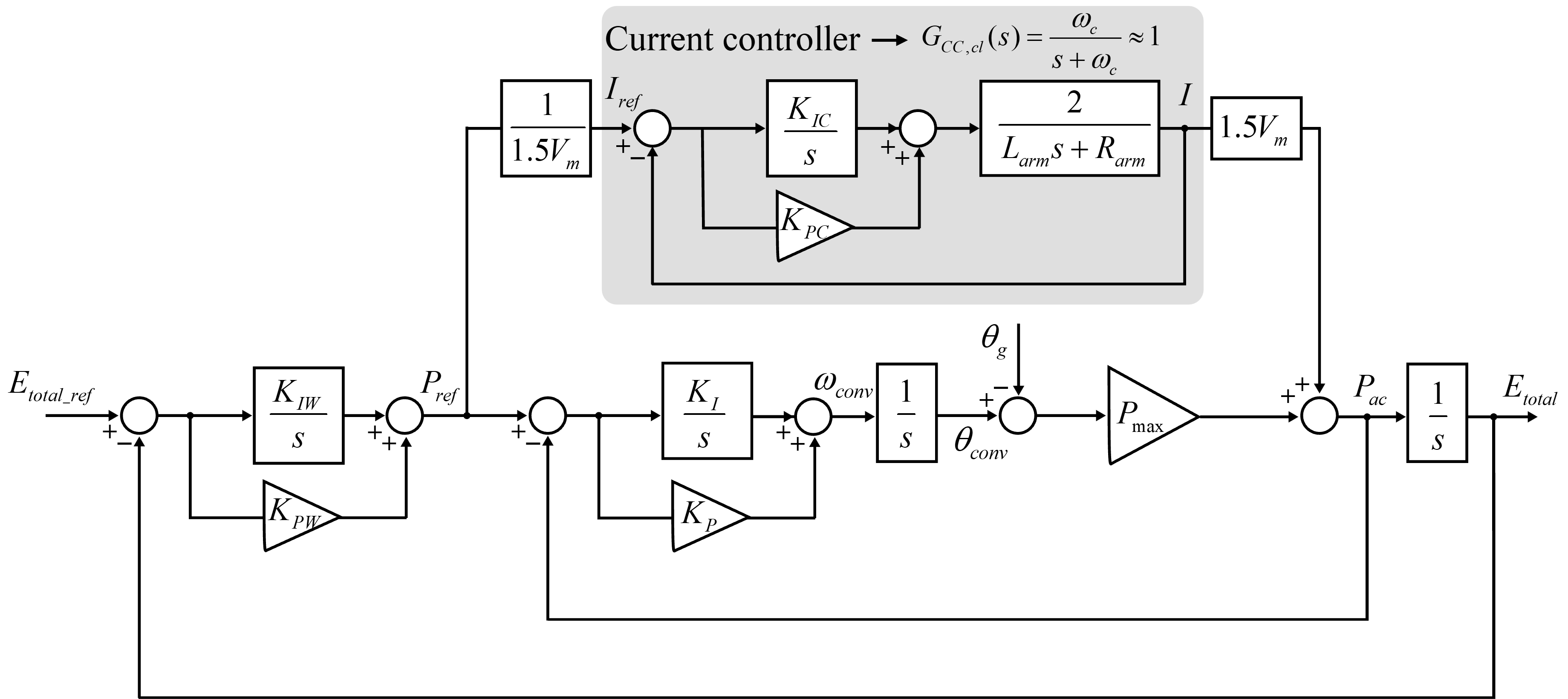}
	\caption{Small-signal model of the cascaded TEC-APC structure with the current-feedforward term.}\label{FIG_7}
\end{figure}

\begin{figure}[!t]\centering
	\includegraphics[width=8.5cm]{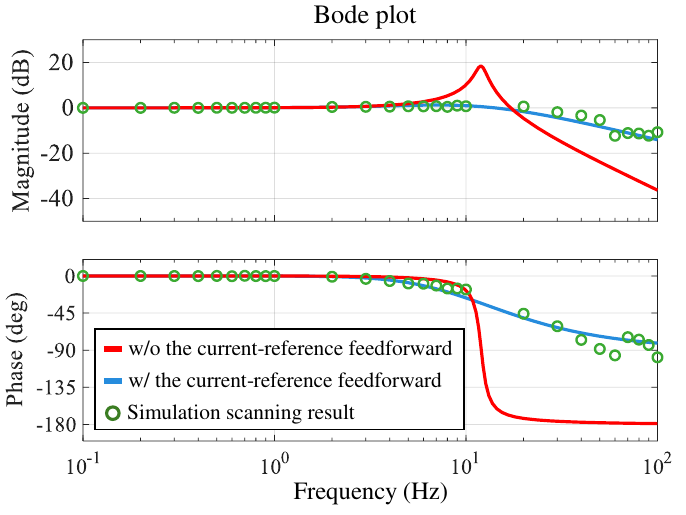}
	\caption{Bode plots of the TEC closed-loop transfer function with and without the current-reference feedforward term.}\label{FIG_8}
\end{figure}

Fig. \ref{FIG_7} presents the small-signal closed-loop model of the cascaded TEC-APC structure with the current-feedforward term.
In general, the closed-loop transfer function of the current controller can be approximated by a first-order low-pass filter \cite{24}.
However, since the current-control bandwidth is much higher than those of the TEC and APC, it can be approximated as unity for the subsequent analysis.
The closed-loop transfer function of the active-power loop including the current-feedforward term is given by (5).
Accordingly, with the current-feedforward term, the slow APC dynamics can be effectively decoupled from the TEC loop.

\begin{align}
G_{\mathrm{APC,cl}}(s) = \frac{P_{ac}}{P_{ref}} = \frac{1 + (K_{P} + \frac{K_I}{s})\frac{P_{max}}{s}}{G_{CC,cl}(s) + (K_{P} + \frac{K_I}{s})\frac{P_{max}}{s}} \approx 1.
\end{align}

Fig. \ref{FIG_8} shows the Bode plots of the TEC closed-loop transfer function with and without the current-reference feedforward term.
The results confirm that the current-feedforward term effectively suppresses the magnitude peaking around 10 - 20 Hz.
Based on the offline simulation scanning results, it is confirmed that, even with the cascaded TEC-APC structure, the slow APC dynamics are effectively decoupled from the TEC loop, consistent with the derived model.

\subsection{Derivation of the Inertial-Response Power-Reference Mapping Term}

As discussed in the previous section, when the total internal energy is regulated through the ac-side power, the TEC output and the inertial-response component appear superimposed in the PCC active power.
Consequently, if the dc-side ES is not used to supply the inertial response, the total internal energy regulation objective constrains the available energy balance and may significantly limit—or even suppress—the desired inertial active-power injection or absorption during a grid-frequency variation.

\begin{figure}[!t]\centering
	\includegraphics[width=8.5cm]{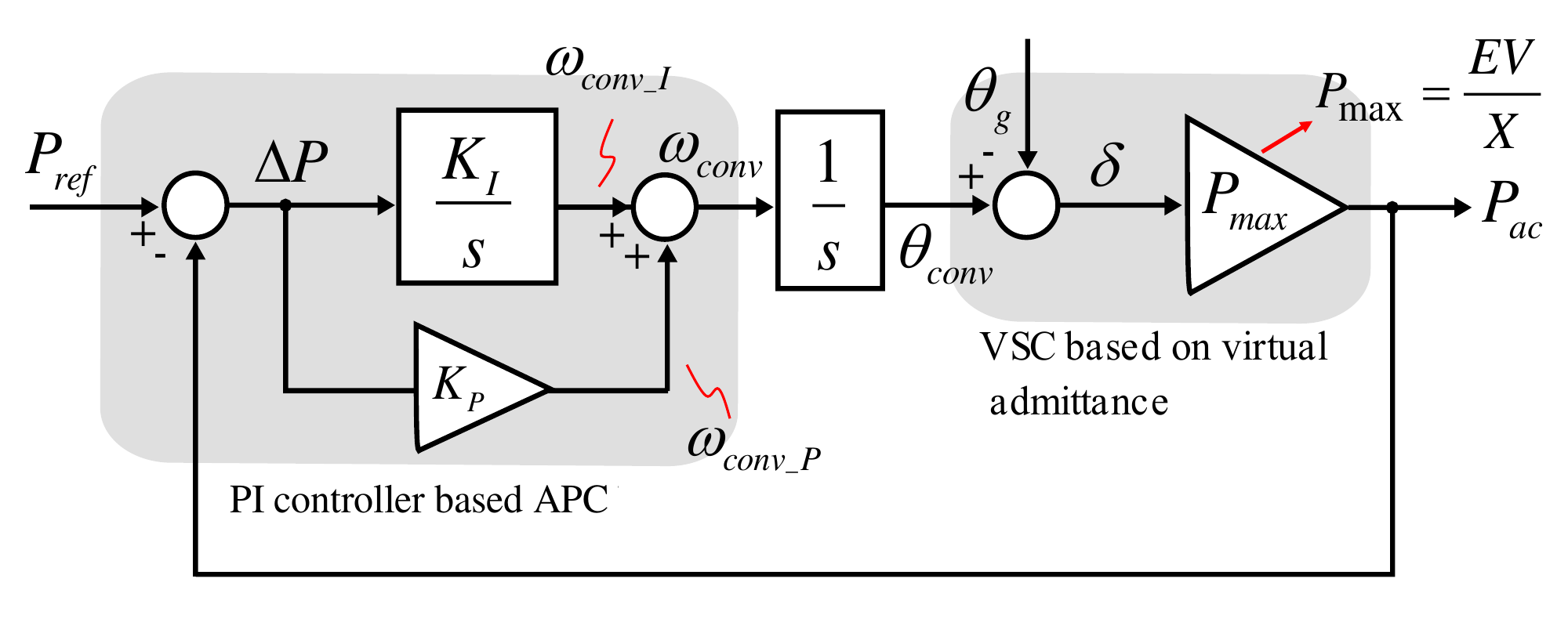}
	\caption{Linearized model of the PI-controller-based APC.}\label{FIG_9}
\end{figure}

Since the overall E-STATCOM controller is implemented within a single main digital signal processor (DSP), the inertial-response power-reference for the dc-side power can be generated by the internal APC.
Fig. \ref{FIG_9} shows the linearized model of the PI-controller-based APC.
The active-power response to a grid-frequency variation is expressed by (6), which is derived from Fig. \ref{FIG_9}.

\begin{align}
\frac{P_{ac}}{\omega_{g}} \approx \frac{-P_{max}s}{s^2 + 2\zeta\omega_n s + \omega_n^2} .
\end{align}

Since the resulting active power is proportional to the time derivative of the grid frequency deviation, (6) exhibits an inertial-response characteristic.
In addition, the converter-frequency response to the grid frequency variation is given by (7).

\begin{align}
\frac{\omega_{conv}}{\omega_{g}} \approx \frac{-K_{P}P_{max}s-K_{I}P_{max}}{s^2 + 2\zeta\omega_n s + \omega_n^2} .
\end{align}

The converter frequency \(\omega_{conv}\) can be decomposed into two components corresponding to the proportional and integral gains of the APC:

\begin{align}
\omega_{conv} = \omega_{conv\_P} + \omega_{conv\_I} .
\end{align}

where \(\omega_{conv\_P}\) and \(\omega_{conv\_I}\) represent the contributions from the proportional and integral terms, respectively.
The proportional component \(\omega_{conv\_P}\) exhibits a response to grid frequency deviations that is proportional to the rate of change of frequency (i.e., \(d\omega_{g}/dt\)), which is characteristic of inertial response. 
This component is scaled by the proportional gain \(K_P\), as shown in equation (9).


\begin{align}
\frac{\omega_{conv\_P}}{\omega_{g}} \approx \frac{-K_{P}P_{max}s}{s^2 + 2\zeta\omega_n s + \omega_n^2} .
\end{align}

As shown in Fig. \ref{FIG_6}, a deadband can be introduced to prevent the ES output from responding excessively to small converter-frequency variations.
The resulting mismatch caused by the deadband can be effectively mitigated by the APC.
Consequently, the computed inertial-response power reference is combined with the active-power reference used to regulate the ES state of charge (SOC), and the sum is applied to the dc-side power control.
In addition, the proposed E-STATCOM control strategy enables STATCOM-mode operation by bypassing the inertial-power reference, without switching to grid-following (GFL) control to avoid instability in the cascaded TEC-APC structure.

\section{Simluation and Experimental Resutls}

\subsection{Full-scaled simulation results}

\begin{table}[!t]
	\renewcommand{\arraystretch}{1.3}
	\caption{Simulation Parameters}
	\centering
	\label{table_1}
	\resizebox{\columnwidth}{!}{
		\begin{tabular}{l l l}
			\hline\hline \\[-3mm]
			\multicolumn{1}{c}{\pbox{20cm}{Parameters}} & \multicolumn{1}{c}{\pbox{20cm}{Values}}  \\[1ex] \hline
			Rated grid line-to-line rms voltage & 132 kV \\
			Tranformation ratio of delta-wye transformer & 2.7:1 \\
			Rated power of the E-STATCOM system  & 50 MVA \\
			Inductance of arm inductor & 25.46 mH \\
			Capacitance of submodule capacitor & 2.65 mF \\ 
			Rated dc voltage of each submodule & 2 kV \\ 
			Number of full-bridge submodules in each arm & 40 EA \\ 
			Energy storage output DC voltage & 60 kV \\[1ex]
			\hline\hline
		\end{tabular}
	}
\end{table}

\begin{figure*}[!t]
	\centering
	\subfloat[]{\includegraphics[width=0.25\textwidth]{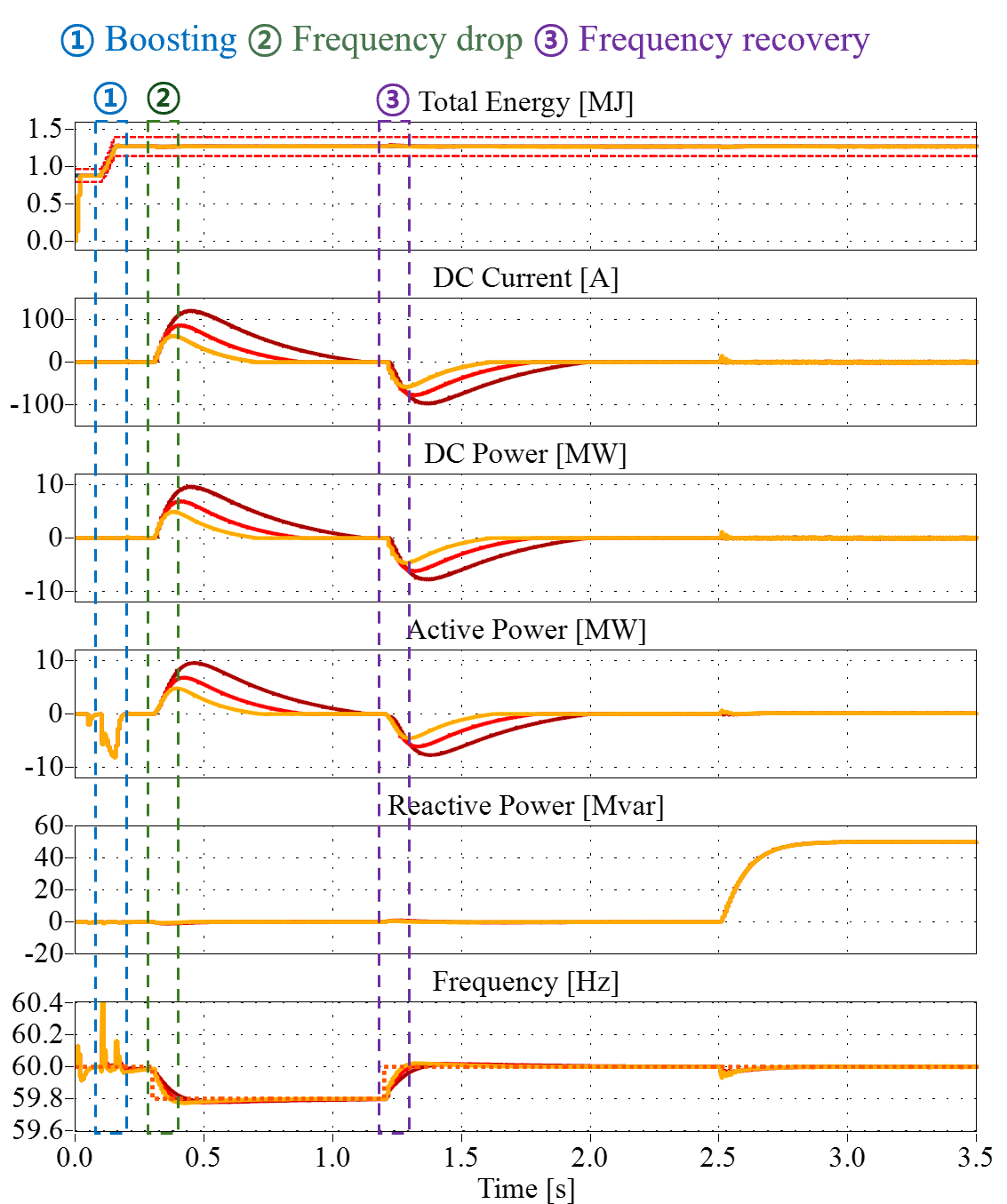}\label{fig:10a}}\hfil
	\subfloat[]{\includegraphics[width=0.25\textwidth]{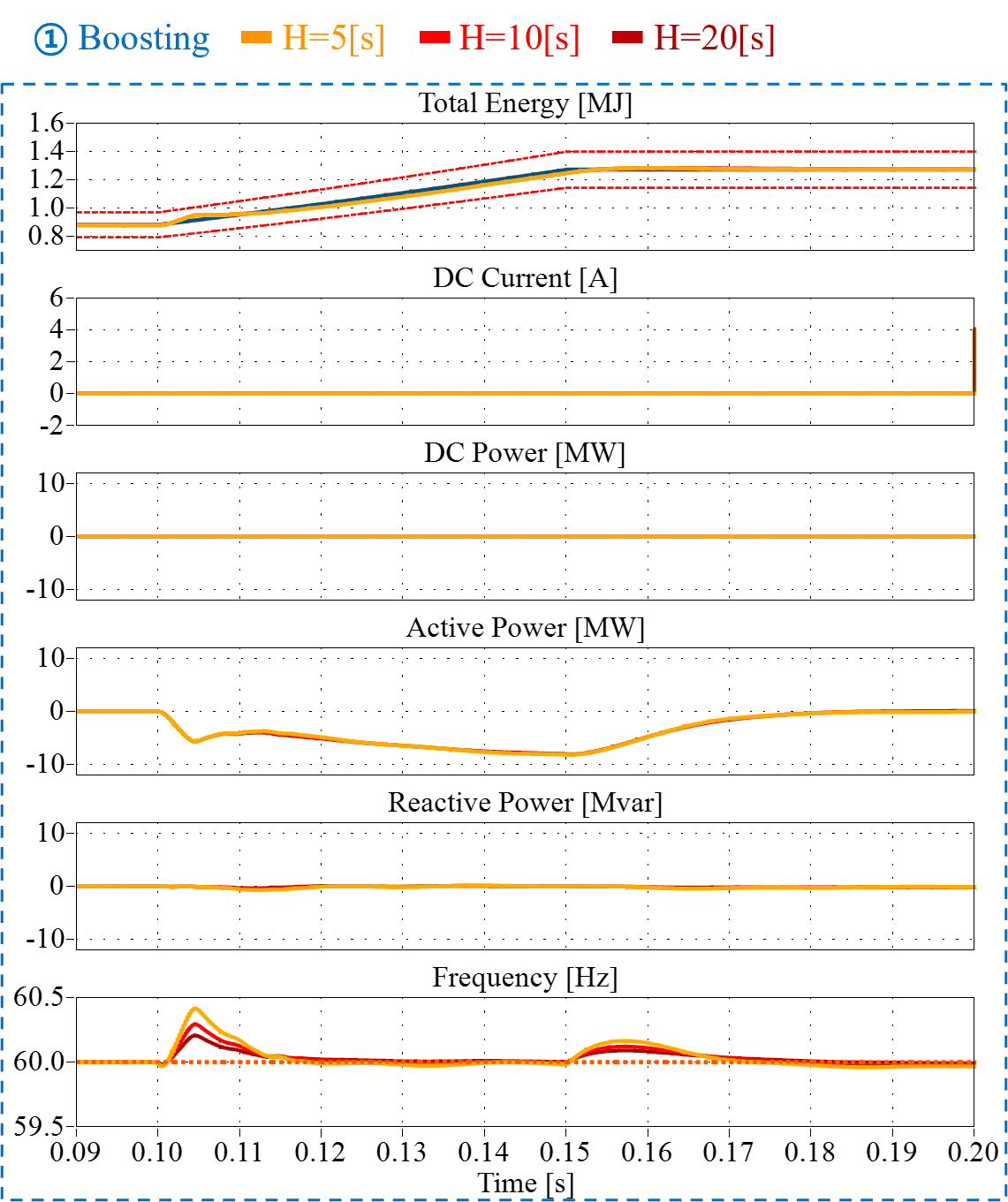}\label{fig:10b}}\hfil
	\subfloat[]{\includegraphics[width=0.25\textwidth]{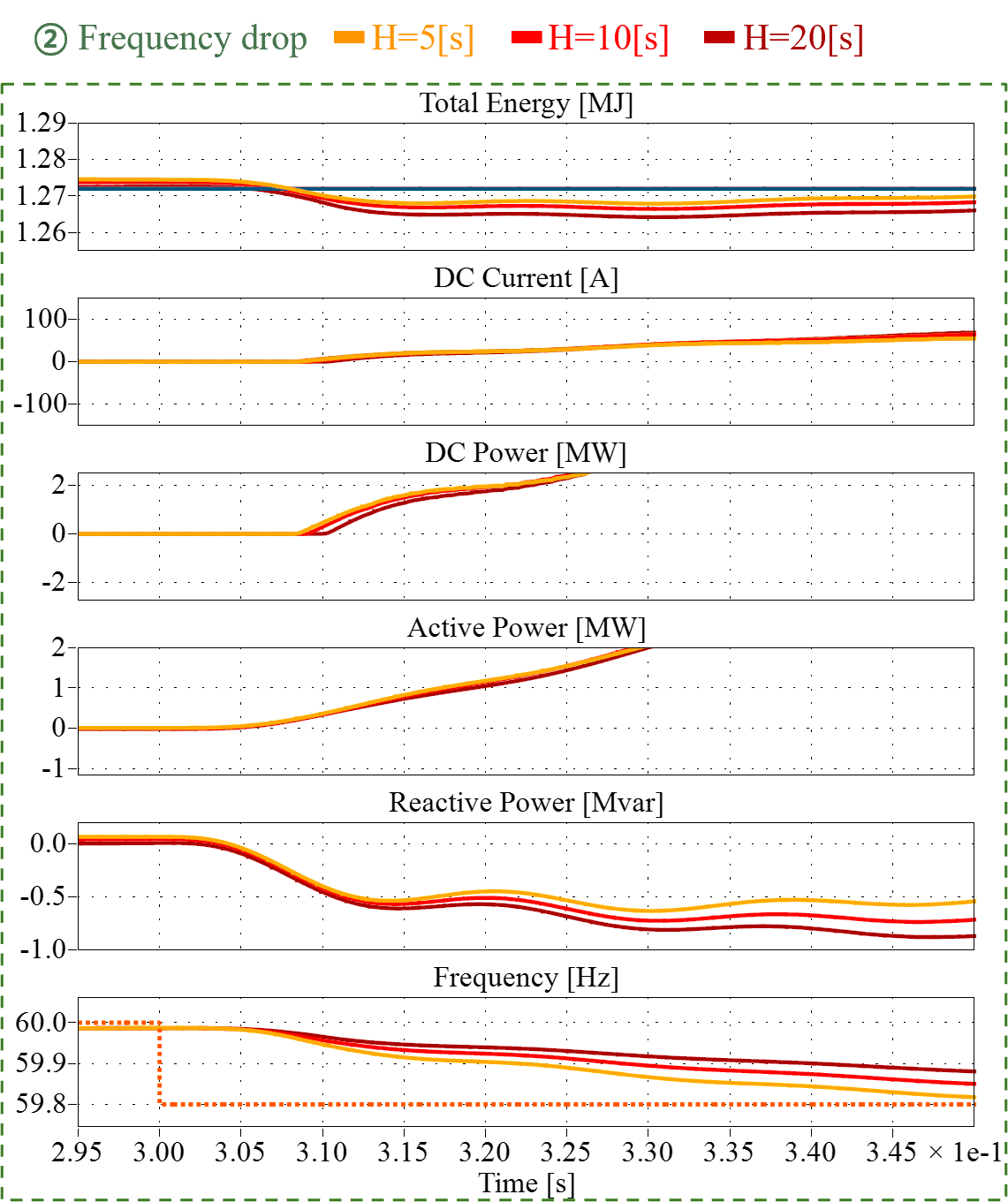}\label{fig:10c}}\hfil
	\subfloat[]{\includegraphics[width=0.25\textwidth]{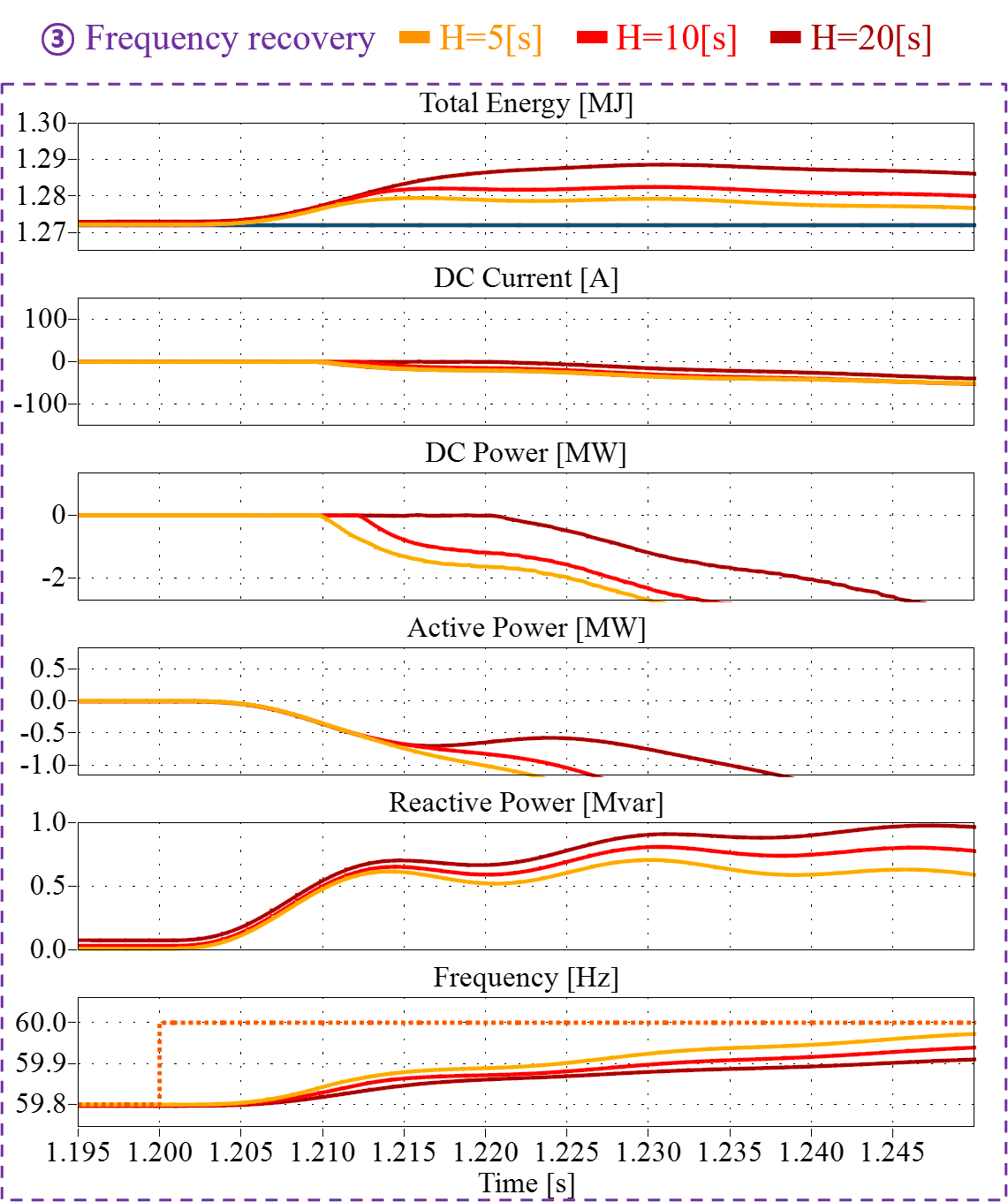}\label{fig:10d}}
	\caption{Simulation results of the proposed control strategy. (a) Overall waveforms for the full simulation scenario. (b) Zoomed-in view of region 1 highlighting total internal energy boosting (blue dashed box). (c) Zoomed-in view of region 2 in which the grid frequency is reduced to 59.8 Hz (green dashed box). (d) Zoomed-in view of region 3 in which the grid frequency is restored to 60 Hz (purple dashed box).}\label{FIG_10}
\end{figure*}

To validate the proposed E-STATCOM control strategy, as shown in Fig. \ref{FIG_6}, a 50MVA offline simulation has been conducted.
Table \ref{table_1} presents the parameters used in the offline simulation.
The simulation scenario is as follows.
From \(t\) = 0 s, passive pre-charge is initiated, and the pre-charge resistor is bypassed at \(t\) = 0.05 s.
At \(t\) = 0.1 s, the total internal-energy boosting and arm-energy balancing controls are activated.
After the boosting stage, the dc magnetic contactor (MC) is closed at \(t\) = 0.2 s.
To evaluate the inertial response, the grid frequency is reduced to 59.8 Hz at \(t\) = 0.3 s and restored to 60 Hz at \(t\) = 1.2 s.
Finally, to verify STATCOM operation, the E-STATCOM is commanded to inject rated reactive power at \(t\) = 2.5 s.

Fig. \ref{FIG_10} presents the offline simulation results.
Figs. \ref{FIG_10}\subref{fig:10b}–\subref{fig:10d} provide zoomed-in views of regions 1–3 in Fig. \ref{FIG_10}\subref{fig:10a}, respectively.
As shown in Fig. \ref{FIG_10}\subref{fig:10b}, the proposed control strategy enables total internal energy boosting through the ac-side active power even when the dc MC remains open.
As shown in Fig. \ref{FIG_10}\subref{fig:10c}, during the grid-frequency decrease, the converter frequency is maintained for a short duration and the inertial response is effectively injected into the grid through the dc-side power.
Similarly, Fig. \ref{FIG_10}\subref{fig:10d} confirms that inertial response is properly absorbed through the dc-side power when the grid frequency is restored.
As the inertial time constant \(H\) increases, the inertial response magnitude increases.
Since a larger \(H\) corresponds to a lower APC bandwidth, the dc-side power exhibits a delayed onset due to the deadband, resulting in a longer non-responsive period.
Consequently, a longer non-responsive period of the dc-side power leads to a larger variation in the total internal energy.
Throughout all scenarios, the total internal energy remains within the $\pm 10 \%$ band indicated by the red dashed lines, and reactive-power injection is properly achieved.
These results verify stable operation of the DS-MC-based E-STATCOM under the proposed strategy.

\subsection{Down-scaled experimental results}

\begin{table}[!t]
	\renewcommand{\arraystretch}{1.3}
	\caption{Experimental Parameters}
	\centering
	\label{table_2}
	\resizebox{\columnwidth}{!}{
		\begin{tabular}{l l l}
			\hline\hline \\[-3mm]
			\multicolumn{1}{c}{\pbox{20cm}{Parameters}} & \multicolumn{1}{c}{\pbox{20cm}{Values}}  \\[1ex] \hline
			Rated grid line-to-line rms voltage & 110 V \\
			Rated power of the E-STATCOM system  & 4 kVA \\
			Inductance of arm inductor & 4 mH \\
			Capacitance of submodule capacitor & 5.4 mF \\ 
			Rated dc voltage of each submodule & 50 V \\ 
			Number of full-bridge submodules in each arm & 5 EA \\ 
			Energy storage output DC voltage & 250 kV \\[1ex]
			\hline\hline
		\end{tabular}
	}
\end{table}

\begin{figure}[!t]\centering
	\includegraphics[width=8.5cm]{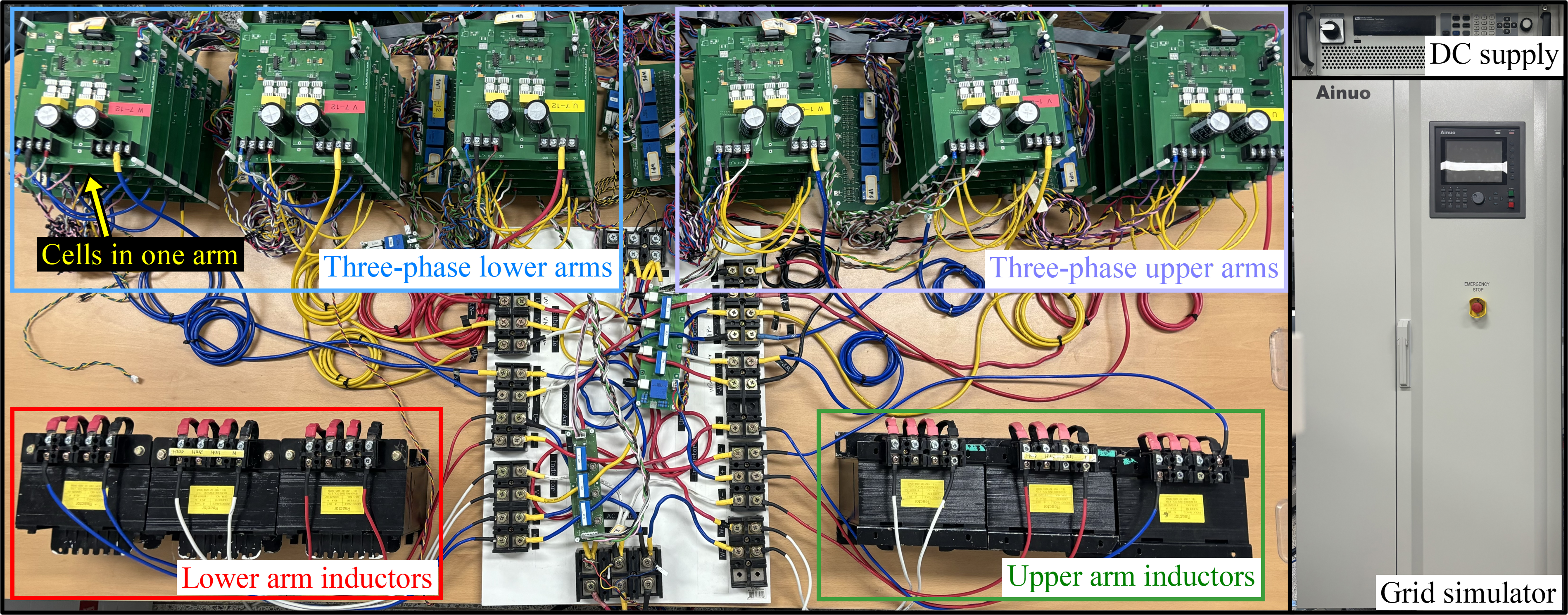}
	\caption{Down-scaled experimental setup.}\label{FIG_11}
\end{figure}

\begin{figure}[!t]
	\centering
	\subfloat[]{\includegraphics[width=8.5cm]{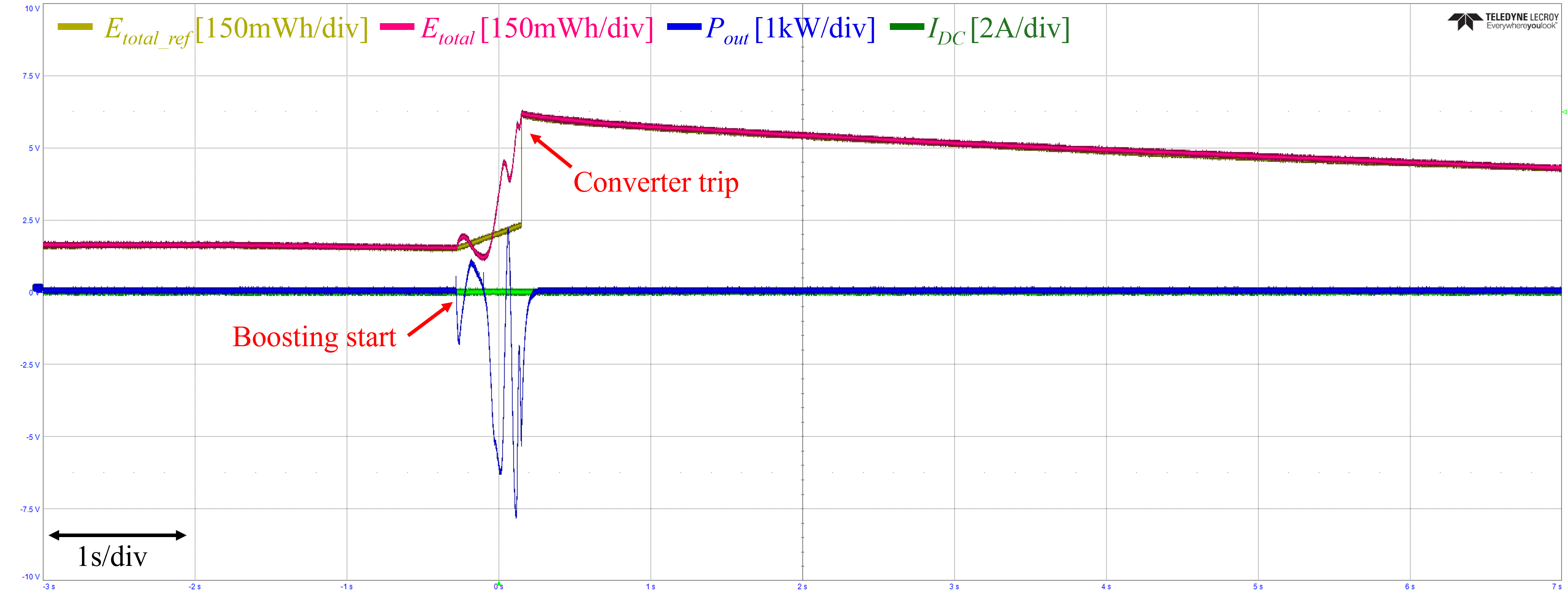}\label{fig:12a}}\\[1mm]
	\subfloat[]{\includegraphics[width=8.5cm]{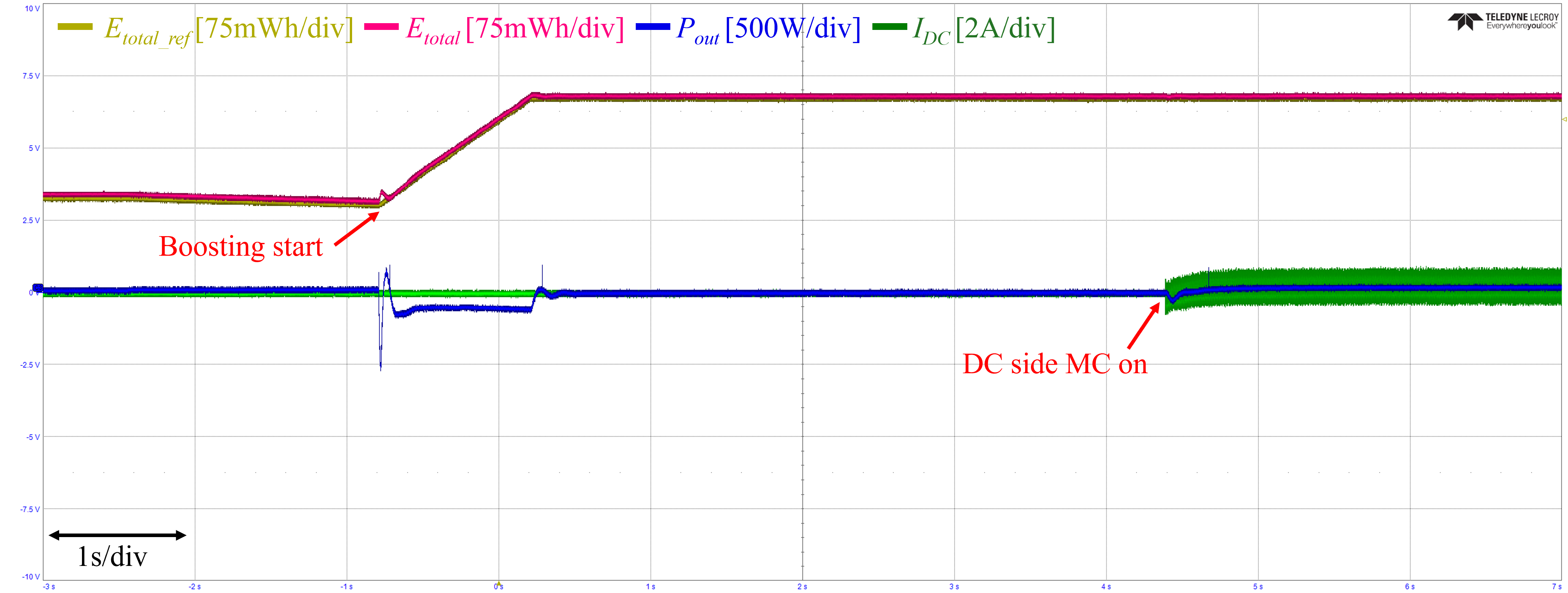}\label{fig:12b}}
	\caption{Experimental results of total internal energy boosting via ac-side active power. (a) without the proposed control strategy. (b) with the proposed control strategy.}\label{FIG_12}
\end{figure}

\begin{figure}[!t]
	\centering
	\subfloat[]{\includegraphics[width=8.5cm]{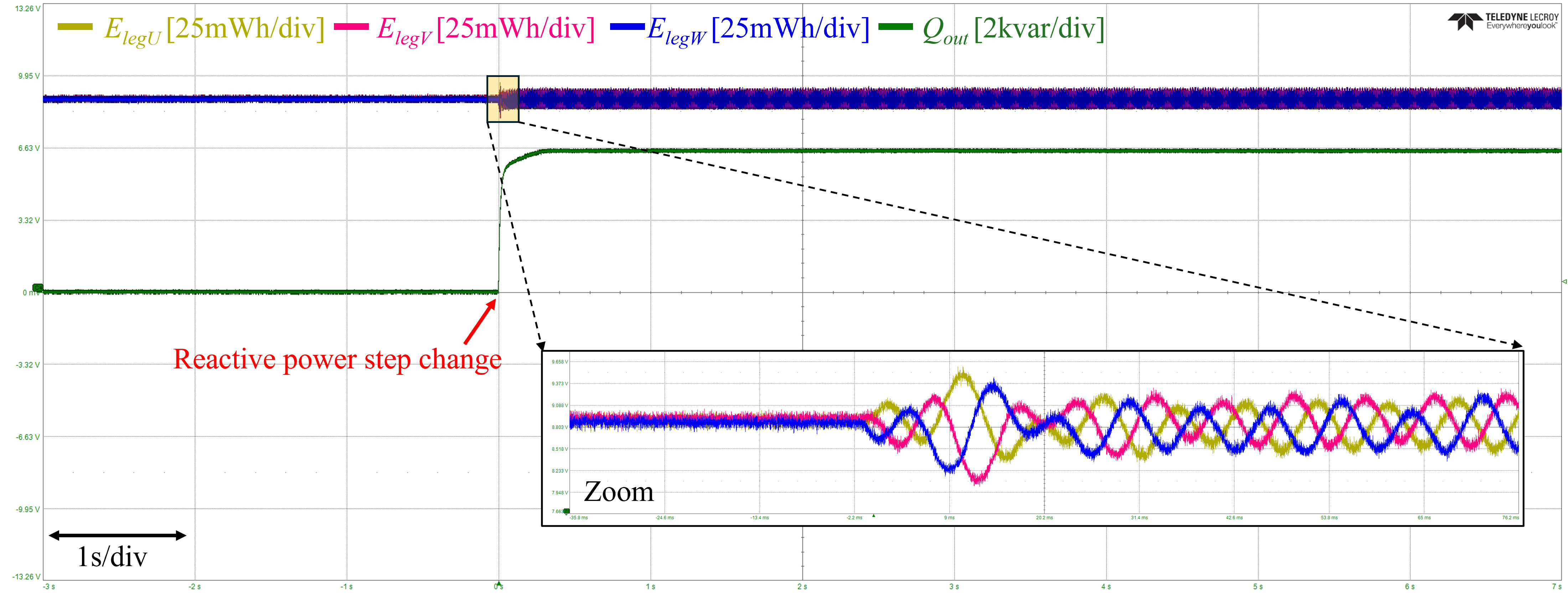}\label{fig:13a}}\\[1mm]
	\subfloat[]{\includegraphics[width=8.5cm]{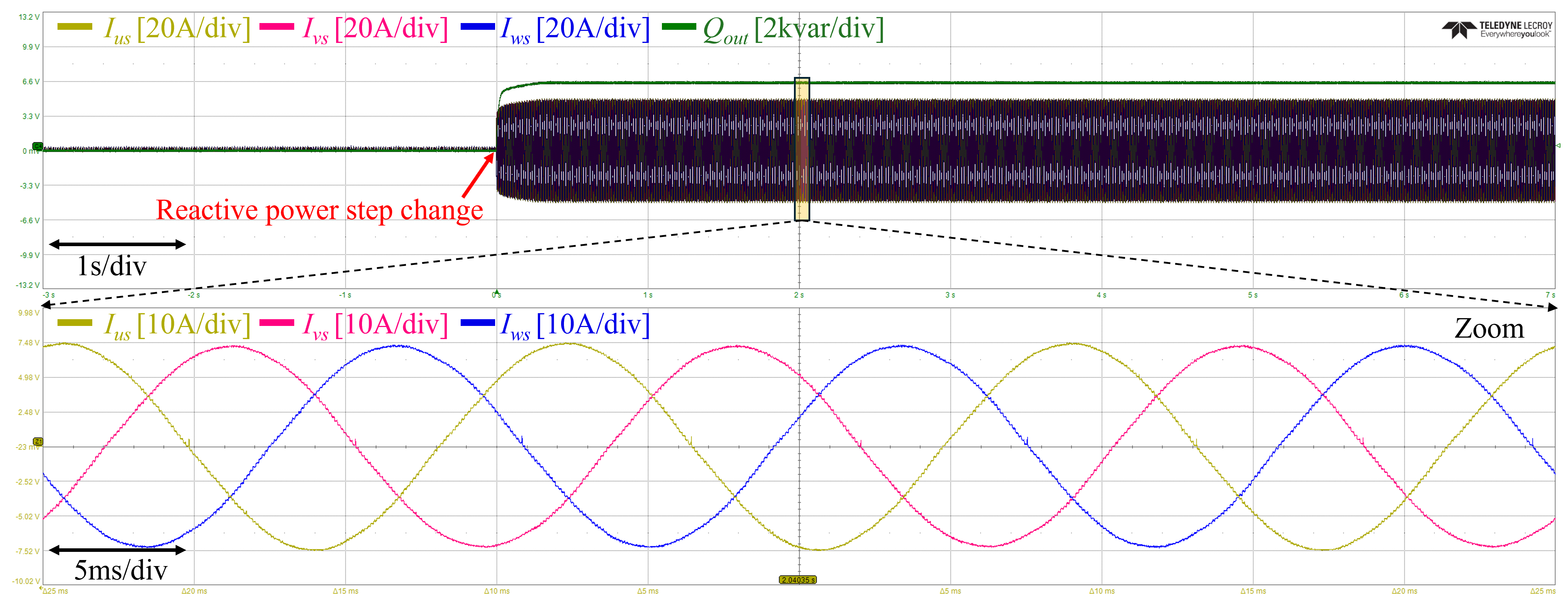}\label{fig:13b}}
	\caption{Experimental results of reactive power injection (a) Leg energies and reactive power. (b) Output current and reactive power.}\label{FIG_13}
\end{figure}

\begin{figure}[!t]
	\centering
	\subfloat[]{\includegraphics[width=4.25cm]{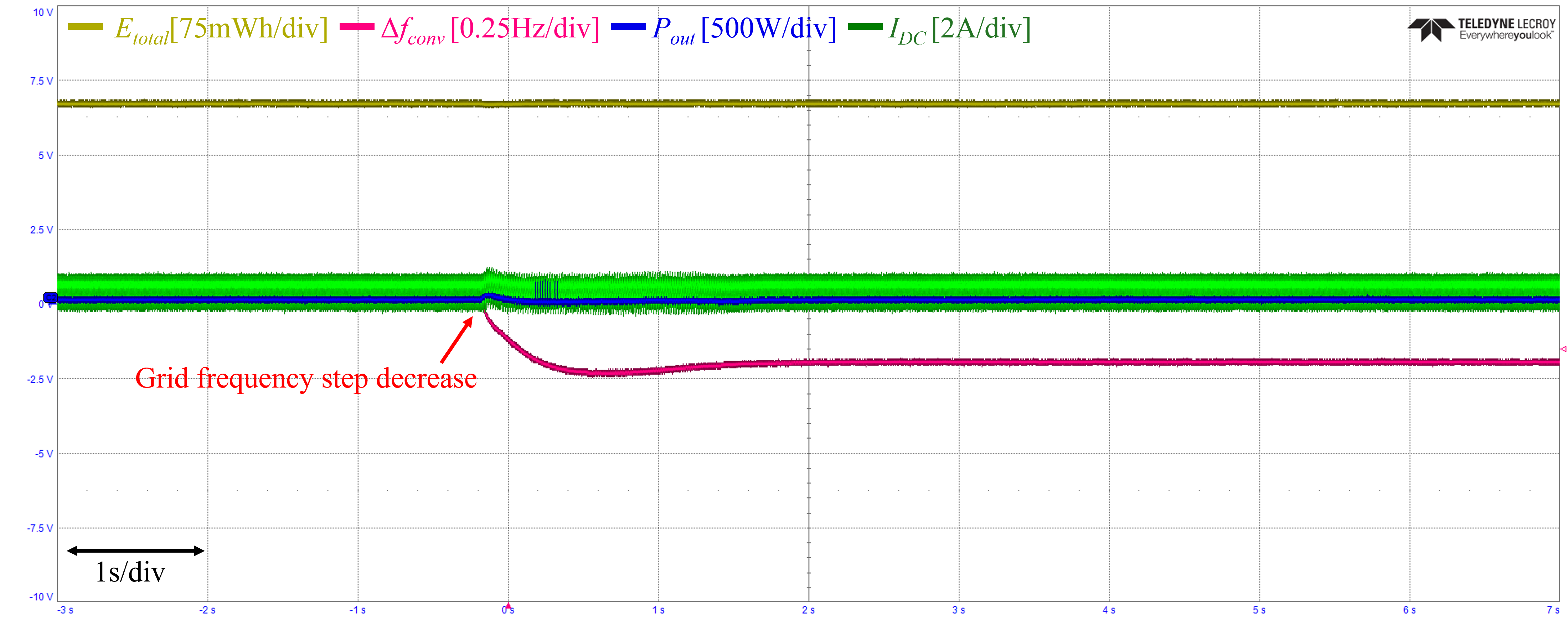}\label{fig:14a}}
	\subfloat[]{\includegraphics[width=4.25cm]{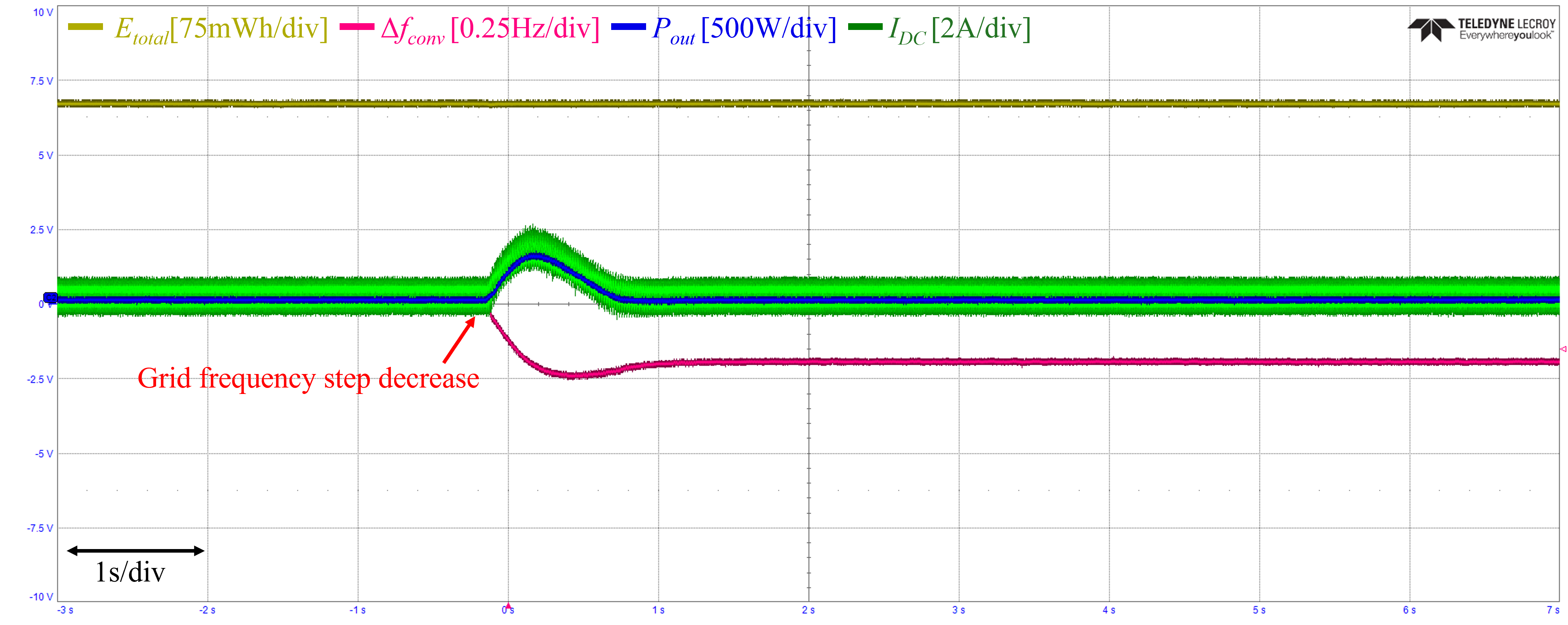}\label{fig:14b}}\\[1mm]
	\subfloat[]{\includegraphics[width=4.25cm]{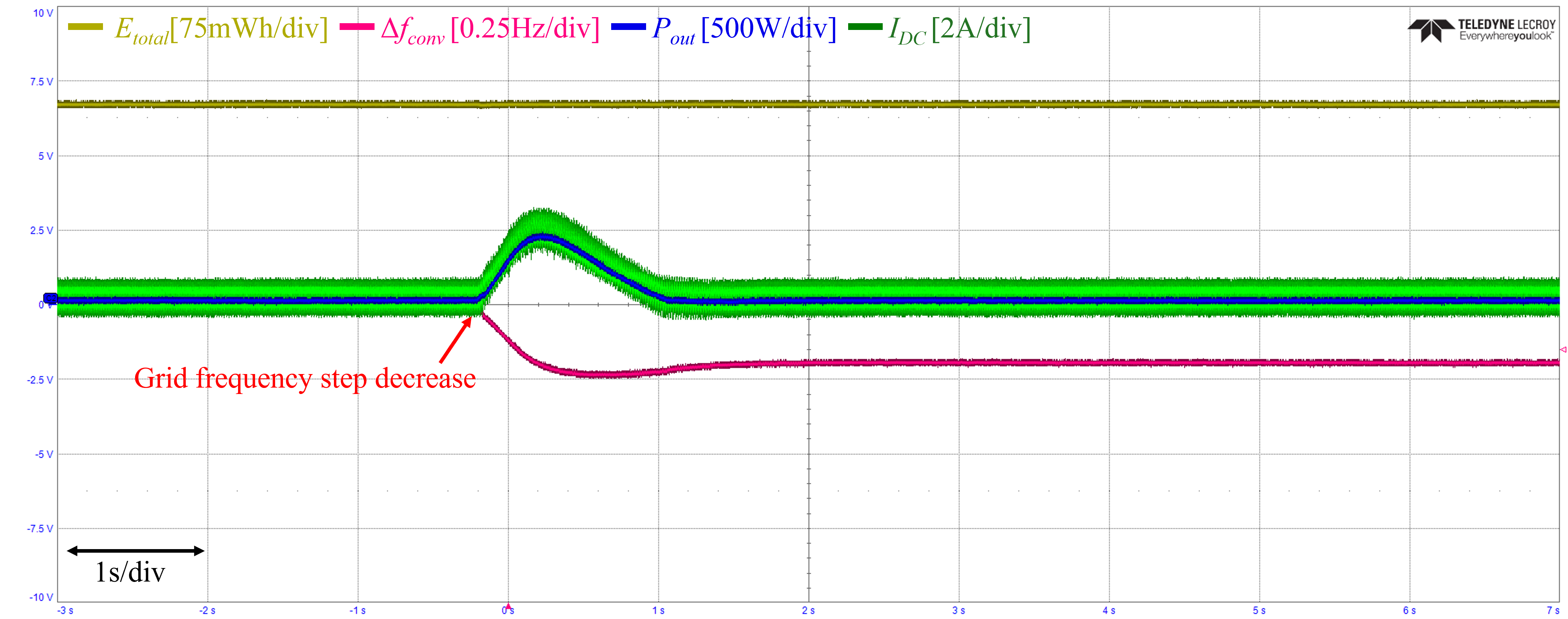}\label{fig:14c}}
	\subfloat[]{\includegraphics[width=4.25cm]{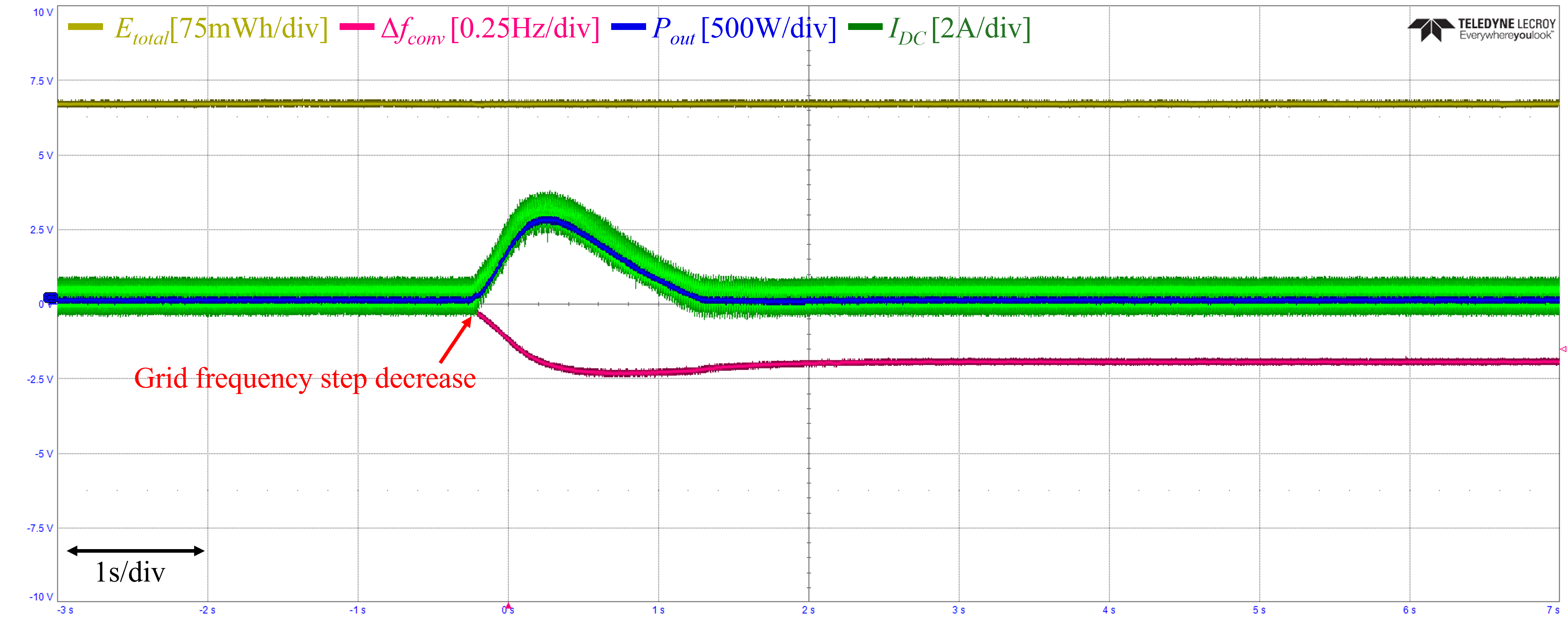}\label{fig:14d}}
	\caption{Experimental results of inertial response for different inertial time constants \(H\). (a) STATCOM mode. (b) \(H\) = 10 s. (c) \(H\) = 20 s. (d) \(H\) = 30 s.}\label{FIG_14}
\end{figure}

\begin{figure}[!t]\centering
	\includegraphics[width=8.5cm]{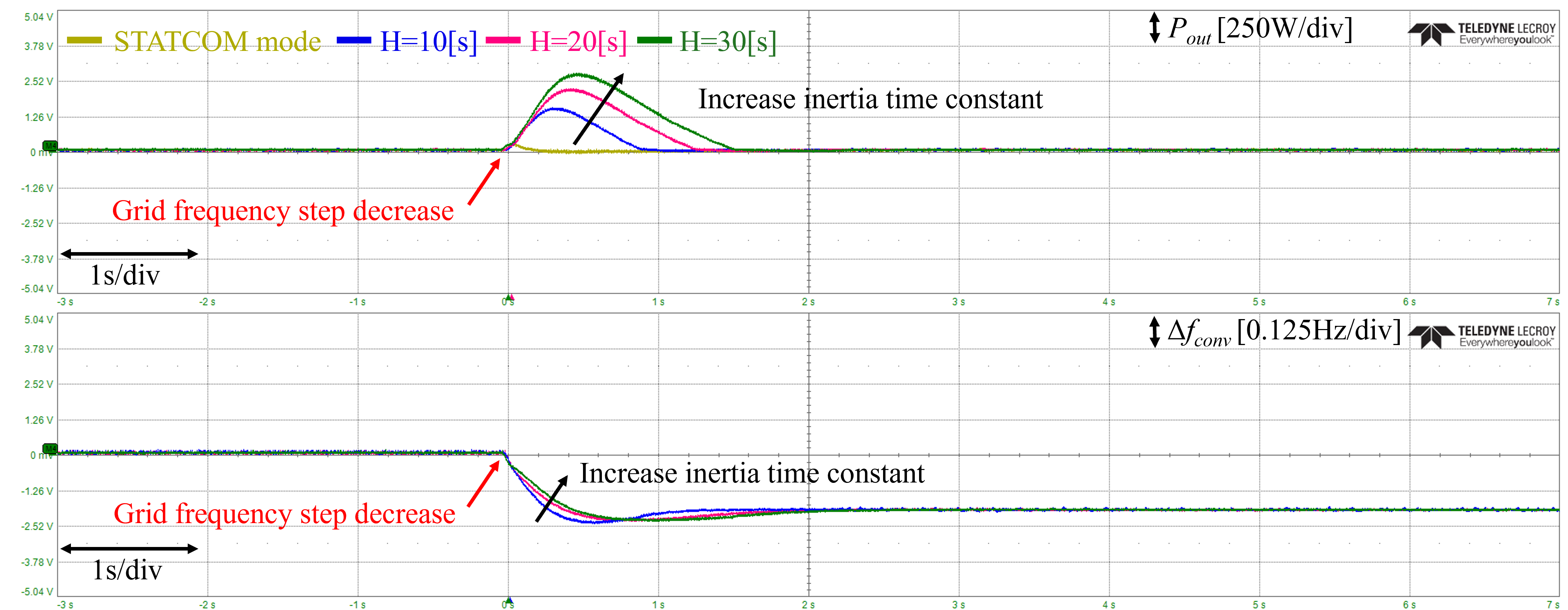}
	\caption{Comparison of inertial response for different inertial time constants \(H\).}\label{FIG_15}
\end{figure}

The proposed E-STATCOM control strategy is validated through a down-scaled experiment setup, as shown in Fig. \ref{FIG_11}, with the parameters listed in Table \ref{table_2}.
Fig. \ref{FIG_12} compares the ac-side active power based boosting performance with and without the proposed E-STATCOM control strategy.
As shown in Fig. \ref{FIG_12}\subref{fig:12a}, without the proposed strategy, improper bandwidth separation causes the DS-MC to trip.
In contrast, Fig. \ref{FIG_12}\subref{fig:12b} confirms that, with the proposed strategy, the total internal energy is properly regulated through the ac-side active power even when the dc MC remains open.

Fig. \ref{FIG_13} presents the experimental results for rated reactive power injection. 
When the E-STATCOM injects the rated reactive power, the leg energies of each phase remain well balanced, while low-THD output current is injected into the grid.

Fig. \ref{FIG_14} shows the inertial response of the E-STATCOM during a grid-frequency decrease for different inertial time constants \(H\).
As shown in Fig. \ref{FIG_14}\subref{fig:14a}, the proposed control strategy enables the E-STATCOM to operate in the STATCOM mode.
Therefore, in the absence of the ES, the E-STATCOM does not deliver active power even when the grid frequency decreases.
Moreover, as shown in Fig. \ref{FIG_14}\subref{fig:14b}-\subref{fig:14d}, the E-STATCOM injects active power into the grid according to the specified inertial time constant during the frequency drop.
Since the inertial-response is allocated to the ES via the dc-side power, the total internal energy of the E-STATCOM is properly maintained throughout the event.

Fig. \ref{FIG_15} compares the inertial responses for different inertial time constants \(H\).
For the same magnitude of frequency decrease, a larger \(H\) yields greater inertial energy, quantified as the time integral of the injected active power.
In addition, since a larger \(H\) implies a lower APC bandwidth, the converter frequency decreases more slowly as \(H\) increases.

\section{Conclusion}

This paper presented a DS-MC-based E-STATCOM control framework that separates internal-energy management from frequency-support functionality.
By regulating the total internal energy through the ac-side active power, the proposed approach avoids continuous loss compensation by the dc-side energy storage, thereby reducing unnecessary energy-throughput demands.
The dc-side ES is consequently reserved for inertial response, while the E-STATCOM retains the ability to operate seamlessly as a conventional STATCOM without mode switching when the storage is unavailable.
Offline simulations and laboratory-scale experiments corroborated that (i) internal-energy boosting can be performed safely before dc MC closure, (ii) rated reactive-power injection is achieved with balanced leg energies and low current distortion, and (iii) inertial response can be shaped via the inertial time constant \(H\) without compromising internal-energy regulation.
These results indicate that the proposed control strategy provides a practical and flexible operating basis for DS-MC-based E-STATCOMs in low-inertia grids.



\bibliographystyle{Bibliography/IEEEtranTIE}
\bibliography{Bibliography/IEEEabrv,Bibliography/BIB_xx-TIE-xxxx}\ 
	
\vspace{-1cm}
\begin{IEEEbiography}[{\includegraphics[width=1in,height=1.25in,clip,keepaspectratio]{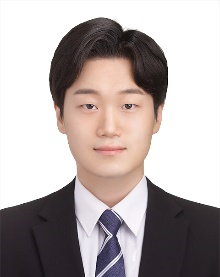}}]
{Ki-Hyun Kim} (Graduate Student Member, IEEE) received a B.S. degree in electronics engineering from Kyungpook National University, Daegu, South Korea, in 2022. 
He is currently pursuing the Ph.D. degree in electronic and electrical engineering from Kyungpook National University. 
His current research interests include grid-connected power electronics, grid-forming converter, and power system dynamics.
\end{IEEEbiography}

\vspace{-1cm}
\begin{IEEEbiography}[{\includegraphics[width=1in,height=1.25in,clip,keepaspectratio]{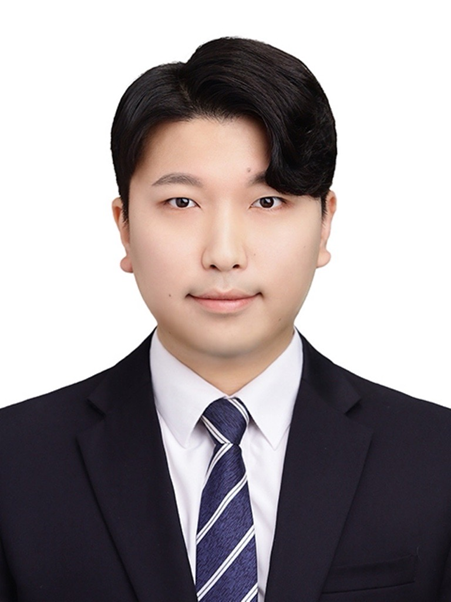}}]
{Yeongung Kim} (Graduate Student Member, IEEE) received the B.S. degree in electrical engineering from Kyungpook National University,Daegu, South Korea, in 2022, where he is currentlypursuing the Ph.D. degree in electronic and electrical engineering. 
His current research interests include the LVDC, MVDC, modular multi-level converter, and microgrid.
\end{IEEEbiography}

\vspace{-1cm}
\begin{IEEEbiography}[{\includegraphics[width=1in,height=1.25in,clip,keepaspectratio]{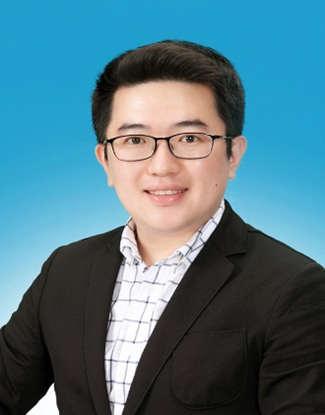}}]
{Shenghui Cui} (Senior Member, IEEE) received the B.S. degree from Tsinghua University, Beijing, China, in 2012, the M.S. degree from Seoul National University, Seoul, South Korea, in 2014, and the Dr.-Ing. Degree with the highest distinction (summa cum laude) from RWTH Aachen University, Aachen, Germany, in 2019, all in electrical engineering.

Since September 2021, Dr. Cui is with Department of Electrical and Computer Engineering, Seoul National University, Seoul, South Kore, where he is currently an associate professor. 
\end{IEEEbiography}

\vspace{-1cm}
\begin{IEEEbiography}[{\includegraphics[width=1in,height=1.25in,clip,keepaspectratio]{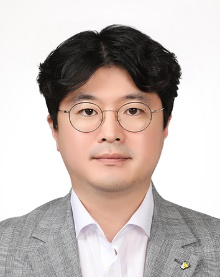}}]
{Jae-Jung Jung} (Senior Member, IEEE) received his B.S., M.S., and Ph.D. degrees in electrical engineering and computer science from Seoul National University, Seoul, South Korea, in 2011, 2013, and 2017, respectively. 
From 2017 to 2019, he was a senior engineer with Samsung Electronics, South Korea. 

Since 2019, he has been a faculty member in the Department of Electrical Engineering at Kyungpook National University in Daegu, South Korea, where he is currently working as an associate professor. 
\end{IEEEbiography}

\end{document}